\documentclass[aps,prl,twocolumn,showpacs,superscriptaddress,nofootinbib,floatfix, longbibliography]{revtex4-2}

\usepackage{amsmath}
\usepackage{graphicx}
\graphicspath{{figures/}}
\usepackage{xspace}
\usepackage{accents}
\usepackage{scalerel}
\usepackage{color}
\usepackage{bm}
\usepackage{subfigure}
\usepackage{alphalph}
\usepackage{hyperref}
\usepackage{multirow}

\begin{document}

\title{Joint neutrino oscillation analysis from the T2K and NOvA experiments}

\collaboration{The T2K and NOvA Collaborations}\noaffiliation

\newcommand{\ANL}{Argonne National Laboratory, Argonne, Illinois 60439, 
USA}
\newcommand{\Bandirma}{Bandirma Onyedi Eyl\"ul University, Faculty of 
Engineering and Natural Sciences, Engineering Sciences Department, 
10200, Bandirma, Balıkesir, Turkey}
\newcommand{\ICS}{Institute of Computer Science, The Czech 
Academy of Sciences, 
182 07 Prague, Czech Republic}
\newcommand{\IOP}{Institute of Physics, The Czech 
Academy of Sciences, 
182 21 Prague, Czech Republic}
\newcommand{\Atlantico}{Universidad del Atlantico,
Carrera 30 No.\ 8-49, Puerto Colombia, Atlantico, Colombia}
\newcommand{\BHU}{Department of Physics, Institute of Science, Banaras 
Hindu University, Varanasi, 221 005, India}
\newcommand{\UCLA}{Physics and Astronomy Department, UCLA, Box 951547, Los 
Angeles, California 90095-1547, USA}
\newcommand{\Caltech}{California Institute of 
Technology, Pasadena, California 91125, USA}
\newcommand{\Cochin}{Department of Physics, Cochin University
of Science and Technology, Kochi 682 022, India}
\newcommand{\Charles}
{Charles University, Faculty of Mathematics and Physics,
 Institute of Particle and Nuclear Physics, Prague, Czech Republic}
\newcommand{\Cincinnati}{Department of Physics, University of Cincinnati, 
Cincinnati, Ohio 45221, USA}
\newcommand{\CSU}{Department of Physics, Colorado 
State University, Fort Collins, CO 80523-1875, USA}
\newcommand{\CTU}{Czech Technical University in Prague,
Brehova 7, 115 19 Prague 1, Czech Republic}
\newcommand{\Dallas}{Physics Department, University of Texas at Dallas,
800 W. Campbell Rd. Richardson, Texas 75083-0688, USA}
\newcommand{\DallasU}{University of Dallas, 1845 E 
Northgate Drive, Irving, Texas 75062 USA}
\newcommand{\Delhi}{Department of Physics and Astrophysics, University of 
Delhi, Delhi 110007, India}
\newcommand{\JINR}{Joint Institute for Nuclear Research,  
Dubna, Moscow region 141980, Russia}
\newcommand{\Erciyes}{
Department of Physics, Erciyes University, Kayseri 38030, Turkey}
\newcommand{\FNAL}{Fermi National Accelerator Laboratory, Batavia, 
Illinois 60510, USA}
\newcommand{\FSU}{Florida State University, Tallahassee, Florida 32306, USA}
\newcommand{\UFG}{Instituto de F\'{i}sica, Universidade Federal de 
Goi\'{a}s, Goi\^{a}nia, Goi\'{a}s, 74690-900, Brazil}
\newcommand{\Guwahati}{Department of Physics, IIT Guwahati, Guwahati, 781 
039, India}
\newcommand{\Harvard}{Department of Physics, Harvard University, 
Cambridge, Massachusetts 02138, USA}
\newcommand{\Houston}{Department of Physics, 
University of Houston, Houston, Texas 77204, USA}
\newcommand{\IHyderabad}{Department of Physics, IIT Hyderabad, Hyderabad, 
502 205, India}
\newcommand{\Hyderabad}{School of Physics, University of Hyderabad, 
Hyderabad, 500 046, India}
\newcommand{\IIT}{Illinois Institute of Technology,
Chicago IL 60616, USA}
\newcommand{\Indiana}{Indiana University, Bloomington, Indiana 47405, 
USA}
\newcommand{\INR}{Institute for Nuclear Research of Russia, Academy of 
Sciences 7a, 60th October Anniversary prospect, Moscow 117312, Russia}
\newcommand{\Iowa}{Department of Physics and Astronomy, Iowa State 
University, Ames, Iowa 50011, USA}
\newcommand{\Irvine}{Department of Physics and Astronomy, 
University of California at Irvine, Irvine, California 92697, USA}
\newcommand{\Jammu}{Department of Physics and Electronics, University of 
Jammu, Jammu Tawi, 180 006, Jammu and Kashmir, India}
\newcommand{\Lebedev}{Nuclear Physics and Astrophysics Division, Lebedev 
Physical 
Institute, Leninsky Prospect 53, 119991 Moscow, Russia}
\newcommand{\Magdalena}{Universidad del Magdalena, Carrera 32 No 22-08 Santa Marta, Colombia}
\newcommand{\MSU}{Department of Physics and Astronomy, Michigan State 
University, East Lansing, Michigan 48824, USA}
\newcommand{\Crookston}{Math, Science and Technology Department, University 
of Minnesota Crookston, Crookston, Minnesota 56716, USA}
\newcommand{\Duluth}{Department of Physics and Astronomy, 
University of Minnesota Duluth, Duluth, Minnesota 55812, USA}
\newcommand{\Minnesota}{School of Physics and Astronomy, University of 
Minnesota Twin Cities, Minneapolis, Minnesota 55455, USA}
\newcommand{\Mississippi}{University of Mississippi, University, Mississippi 38677, USA}
\newcommand{\NISER}{National Institute of Science Education and Research,
Khurda, 752050, Odisha, India}
\newcommand{\Oxford}{Subdepartment of Particle Physics, 
University of Oxford, Oxford OX1 3RH, United Kingdom}
\newcommand{\Panjab}{Department of Physics, Panjab University, 
Chandigarh, 160 014, India}
\newcommand{\Pitt}{Department of Physics, 
University of Pittsburgh, Pittsburgh, Pennsylvania 15260, USA}
\newcommand{\QMU}{Particle Physics Research Centre, 
Department of Physics and Astronomy,
Queen Mary University of London,
London E1 4NS, United Kingdom}
\newcommand{\RAL}{Rutherford Appleton Laboratory, Science 
and 
Technology Facilities Council, Didcot, OX11 0QX, United Kingdom}
\newcommand{\SAlabama}{Department of Physics, University of 
South Alabama, Mobile, Alabama 36688, USA} 
\newcommand{\Carolina}{Department of Physics and Astronomy, University of 
South Carolina, Columbia, South Carolina 29208, USA}
\newcommand{\SDakota}{South Dakota School of Mines and Technology, Rapid 
City, South Dakota 57701, USA}
\newcommand{\SMU}{Department of Physics, Southern Methodist University, 
Dallas, Texas 75275, USA}
\newcommand{\Stanford}{Department of Physics, Stanford University, 
Stanford, California 94305, USA}
\newcommand{\Sussex}{Department of Physics and Astronomy, University of 
Sussex, Falmer, Brighton BN1 9QH, United Kingdom}
\newcommand{\Syracuse}{Department of Physics, Syracuse University,
Syracuse NY 13210, USA}
\newcommand{\Tennessee}{Department of Physics and Astronomy, 
University of Tennessee, Knoxville, Tennessee 37996, USA}
\newcommand{\Texas}{Department of Physics, University of Texas at Austin, 
Austin, Texas 78712, USA}
\newcommand{\Tufts}{Department of Physics and Astronomy, Tufts University, Medford, 
Massachusetts 02155, USA}
\newcommand{\UCL}{Physics and Astronomy Department, University College 
London, 
Gower Street, London WC1E 6BT, United Kingdom}
\newcommand{\Virginia}{Department of Physics, University of Virginia, 
Charlottesville, Virginia 22904, USA}
\newcommand{\OSU}{Ohio State University, Columbus, Ohio 43210, USA}
\newcommand{\WSU}{Department of Mathematics, Statistics, and Physics,
 Wichita State University, 
Wichita, Kansas 67260, USA}
\newcommand{\WandM}{Department of Physics, William \& Mary, 
Williamsburg, Virginia 23187, USA}
\newcommand{\Wisconsin}{Department of Physics, University of 
Wisconsin-Madison, Madison, Wisconsin 53706, USA}
\newcommand{\deceased}{Deceased.}
\affiliation{\Erciyes}
\affiliation{\Atlantico}
\affiliation{\Mississippi}
\affiliation{\FNAL}
\affiliation{\JINR}
\affiliation{\Magdalena}
\affiliation{\Sussex}
\affiliation{\Cincinnati}
\affiliation{\UFG}
\affiliation{\Indiana}
\affiliation{\Iowa}
\affiliation{\Irvine}
\affiliation{\Hyderabad}
\affiliation{\Bandirma}
\affiliation{\Minnesota}
\affiliation{\Panjab}
\affiliation{\Guwahati}
\affiliation{\QMU}
\affiliation{\IHyderabad}
\affiliation{\MSU}
\affiliation{\CSU}
\affiliation{\INR}
\affiliation{\UCL}
\affiliation{\Texas}
\affiliation{\Wisconsin}
\affiliation{\WandM}
\affiliation{\IIT}
\affiliation{\Delhi}
\affiliation{\SMU}
\affiliation{\Virginia}
\affiliation{\ANL}
\affiliation{\Houston}
\affiliation{\WSU}
\affiliation{\IOP}
\affiliation{\SAlabama}
\affiliation{\Tufts}
\affiliation{\Duluth}
\affiliation{\ICS}
\affiliation{\Syracuse}
\affiliation{\Carolina}
\affiliation{\CTU}
\affiliation{\Cochin}
\affiliation{\Caltech}
\affiliation{\Pitt}
\affiliation{\NISER}
\affiliation{\FSU}
\affiliation{\BHU}
\affiliation{\Charles}
\affiliation{\OSU}

\newcommand{\INSTHD}{\affiliation{University Autonoma Madrid, Department of Theoretical Physics, 28049 Madrid, Spain}}
\newcommand{\INSTFE}{\affiliation{Boston University, Department of Physics, Boston, Massachusetts, U.S.A.}}
\newcommand{\INSTD}{\affiliation{University of British Columbia, Department of Physics and Astronomy, Vancouver, British Columbia, Canada}}
\newcommand{\INSTGA}{\affiliation{University of California, Irvine, Department of Physics and Astronomy, Irvine, California, U.S.A.}}
\newcommand{\INSTI}{\affiliation{IRFU, CEA, Universit\'e Paris-Saclay, F-91191 Gif-sur-Yvette, France}}
\newcommand{\INSTGB}{\affiliation{University of Colorado at Boulder, Department of Physics, Boulder, Colorado, U.S.A.}}
\newcommand{\INSTFH}{\affiliation{Duke University, Department of Physics, Durham, North Carolina, U.S.A.}}
\newcommand{\INSTJA}{\affiliation{E\"{o}tv\"{o}s Lor\'{a}nd University, Department of Atomic Physics, Budapest, Hungary}}
\newcommand{\INSTEF}{\affiliation{ETH Zurich, Institute for Particle Physics and Astrophysics, Zurich, Switzerland}}
\newcommand{\INSTIG}{\affiliation{VNU University of Science, Vietnam National University, Hanoi, Vietnam}}
\newcommand{\INSTIE}{\affiliation{CERN European Organization for Nuclear Research, CH-1211 Gen\'eve 23, Switzerland}}
\newcommand{\INSTEG}{\affiliation{University of Geneva, Section de Physique, DPNC, Geneva, Switzerland}}
\newcommand{\INSTHJ}{\affiliation{University of Glasgow, School of Physics and Astronomy, Glasgow, United Kingdom}}
\newcommand{\INSTJG}{\affiliation{Ghent University, Department of Physics and Astronomy, Proeftuinstraat 86, B-9000 Gent, Belgium}}
\newcommand{\INSTDG}{\affiliation{H. Niewodniczanski Institute of Nuclear Physics PAN, Cracow, Poland}}
\newcommand{\INSTCB}{\affiliation{High Energy Accelerator Research Organization (KEK), Tsukuba, Ibaraki, Japan}}
\newcommand{\INSTED}{\affiliation{Institut de Fisica d'Altes Energies (IFAE) - The Barcelona Institute of Science and Technology, Campus UAB, Bellaterra (Barcelona) Spain}}
\newcommand{\INSTJC}{\affiliation{Institut f\"ur Physik, Johannes Gutenberg-Universit\"at Mainz, Staudingerweg 7, 55128 Mainz, Germany}}
\newcommand{\INSTHH}{\affiliation{Institute For Interdisciplinary Research in Science and Education (IFIRSE), ICISE, Quy Nhon, Vietnam}}
\newcommand{\INSTEI}{\affiliation{Imperial College London, Department of Physics, London, United Kingdom}}
\newcommand{\INSTGF}{\affiliation{INFN Sezione di Bari and Universit\`a e Politecnico di Bari, Dipartimento Interuniversitario di Fisica, Bari, Italy}}
\newcommand{\INSTBE}{\affiliation{INFN Sezione di Napoli and Universit\`a di Napoli, Dipartimento di Fisica, Napoli, Italy}}
\newcommand{\INSTBF}{\affiliation{INFN Sezione di Padova and Universit\`a di Padova, Dipartimento di Fisica, Padova, Italy}}
\newcommand{\INSTBD}{\affiliation{INFN Sezione di Roma and Universit\`a di Roma ``La Sapienza'', Roma, Italy}}
\newcommand{\INSTHI}{\affiliation{International Centre of Physics, Institute of Physics (IOP), Vietnam Academy of Science and Technology (VAST), 10 Dao Tan, Ba Dinh, Hanoi, Vietnam}}
\newcommand{\INSTJD}{\affiliation{ILANCE, CNRS – University of Tokyo International Research Laboratory, Kashiwa, Chiba 277-8582, Japan}}
\newcommand{\INSTHA}{\affiliation{Kavli Institute for the Physics and Mathematics of the Universe (WPI), The University of Tokyo Institutes for Advanced Study, University of Tokyo, Kashiwa, Chiba, Japan}}
\newcommand{\INSTID}{\affiliation{Keio University, Department of Physics, Kanagawa, Japan}}
\newcommand{\INSTIF}{\affiliation{King's College London, Department of Physics, Strand, London WC2R 2LS, United Kingdom}}
\newcommand{\INSTCC}{\affiliation{Kobe University, Kobe, Japan}}
\newcommand{\INSTCD}{\affiliation{Kyoto University, Department of Physics, Kyoto, Japan}}
\newcommand{\INSTEJ}{\affiliation{Lancaster University, Physics Department, Lancaster, United Kingdom}}
\newcommand{\INSTII}{\affiliation{Lawrence Berkeley National Laboratory, Berkeley, California, U.S.A.}}
\newcommand{\INSTBA}{\affiliation{Ecole Polytechnique, IN2P3-CNRS, Laboratoire Leprince-Ringuet, Palaiseau, France}}
\newcommand{\INSTFC}{\affiliation{University of Liverpool, Department of Physics, Liverpool, United Kingdom}}
\newcommand{\INSTFI}{\affiliation{Louisiana State University, Department of Physics and Astronomy, Baton Rouge, Louisiana, U.S.A.}}
\newcommand{\INSTCE}{\affiliation{Miyagi University of Education, Department of Physics, Sendai, Japan}}
\newcommand{\INSTDF}{\affiliation{National Centre for Nuclear Research, Warsaw, Poland}}
\newcommand{\INSTFJ}{\affiliation{State University of New York at Stony Brook, Department of Physics and Astronomy, Stony Brook, New York, U.S.A.}}
\newcommand{\INSTEH}{\affiliation{STFC, Rutherford Appleton Laboratory, Harwell Oxford,  and  Daresbury Laboratory, Warrington, United Kingdom}}
\newcommand{\INSTGJ}{\affiliation{Okayama University, Department of Physics, Okayama, Japan}}
\newcommand{\INSTCF}{\affiliation{Osaka Metropolitan University, Department of Physics, Osaka, Japan}}
\newcommand{\INSTGG}{\affiliation{Oxford University, Department of Physics, Oxford, United Kingdom}}
\newcommand{\INSTIC}{\affiliation{University of Pennsylvania, Department of Physics and Astronomy,  Philadelphia, Pennsylvania, U.S.A.}}
\newcommand{\INSTGC}{\affiliation{University of Pittsburgh, Department of Physics and Astronomy, Pittsburgh, Pennsylvania, U.S.A.}}
\newcommand{\INSTGD}{\affiliation{University of Rochester, Department of Physics and Astronomy, Rochester, New York, U.S.A.}}
\newcommand{\INSTHC}{\affiliation{Royal Holloway University of London, Department of Physics, Egham, Surrey, United Kingdom}}
\newcommand{\INSTBC}{\affiliation{RWTH Aachen University, III. Physikalisches Institut, Aachen, Germany}}
\newcommand{\INSTJF}{\affiliation{School of Physics and Astronomy, University of Minnesota, Minneapolis, Minnesota, U.S.A.}}
\newcommand{\INSTFB}{\affiliation{University of Sheffield, School of Mathematical and Physical Sciences, Sheffield, United Kingdom}}
\newcommand{\INSTDI}{\affiliation{University of Silesia, Institute of Physics, Katowice, Poland}}
\newcommand{\INSTIA}{\affiliation{SLAC National Accelerator Laboratory, Stanford University, Menlo Park, California, U.S.A.}}
\newcommand{\INSTBB}{\affiliation{Sorbonne Universit\'e, CNRS/IN2P3, Laboratoire de Physique Nucl\'eaire et de Hautes Energies (LPNHE), Paris, France}}
\newcommand{\INSTJE}{\affiliation{South Dakota School of Mines and Technology, 501 East Saint Joseph Street, Rapid City, SD 57701, United States}}
\newcommand{\INSTCH}{\affiliation{University of Tokyo, Department of Physics, Tokyo, Japan}}
\newcommand{\INSTBJ}{\affiliation{University of Tokyo, Institute for Cosmic Ray Research, Kamioka Observatory, Kamioka, Japan}}
\newcommand{\INSTCG}{\affiliation{University of Tokyo, Institute for Cosmic Ray Research, Research Center for Cosmic Neutrinos, Kashiwa, Japan}}
\newcommand{\INSTHF}{\affiliation{Institute of Science Tokyo, Department of Physics, Tokyo}}
\newcommand{\INSTGI}{\affiliation{Tokyo Metropolitan University, Department of Physics, Tokyo, Japan}}
\newcommand{\INSTHG}{\affiliation{Tokyo University of Science, Faculty of Science and Technology, Department of Physics, Noda, Chiba, Japan}}
\newcommand{\INSTB}{\affiliation{TRIUMF, Vancouver, British Columbia, Canada}}
\newcommand{\INSTJH}{\affiliation{University of Toyama, Department of Physics, Toyama, Japan}}
\newcommand{\INSTDJ}{\affiliation{University of Warsaw, Faculty of Physics, Warsaw, Poland}}
\newcommand{\INSTDH}{\affiliation{Warsaw University of Technology, Institute of Radioelectronics and Multimedia Technology, Warsaw, Poland}}
\newcommand{\INSTIJ}{\affiliation{Tohoku University, Faculty of Science, Department of Physics, Miyagi, Japan}}
\newcommand{\INSTFD}{\affiliation{University of Warwick, Department of Physics, Coventry, United Kingdom}}
\newcommand{\INSTEA}{\affiliation{Wroclaw University, Faculty of Physics and Astronomy, Wroclaw, Poland}}
\newcommand{\INSTHE}{\affiliation{Yokohama National University, Department of Physics, Yokohama, Japan}}
\newcommand{\INSTH}{\affiliation{York University, Department of Physics and Astronomy, Toronto, Ontario, Canada}}
\newcommand{\INSTJB}{\affiliation{Departamento de F\'isica At\'omica, Molecular y Nuclear, Universidad de Sevilla, 41080 Sevilla, Spain}}
\newcommand{\INSTBL}{\affiliation{Departament de Fisica de la Universitat Autonoma de Barcelona, Barcelona, Spain}}
\newcommand{\INSTJP}{\affiliation{Japan Proton Accelerator Research Complex, Tokai, Japan}}
\newcommand{\INSTNB}{\affiliation{Nambu Yoichiro Institute of Theoretical and Experimental Physics, Osaka, Japan}}
\newcommand{\INSTCU}{\affiliation{Science Department, Borough of Manhattan Community College, City Univerisity of New York, New York, New York, U.S.A.}}
\newcommand{\INSTVI}{\affiliation{Graduate University of Science and Technology, Vietnam Academy of Science and Technology, Hanoi, Vietnam}}
\newcommand{\INSTIP}{\affiliation{IPSA-DRII, Ivry-sur-Seine, France}}
\newcommand{\INSTPS}{\affiliation{Universit\'e Paris-Saclay, F-91191 Gif-sur-Yvette, France}}
\newcommand{\INSTWP}{\affiliation{Faculty of Physics, University of Warsaw, Warsaw, Poland}}

{\INSTCG}
{\INSTWP}
{\INSTCB}
{\INSTEJ}
{\INSTFB}
{\INSTEF}
{\INSTEI}
{\INSTCC}
{\INSTCH}
{\INSTGI}
{\INSTCD}
{\INSTIJ}
{\INSTFI}
{\INSTFD}
{\INSTGG}
{\INSTFC}
{\INSTDG}
{\INSTGF}
{\INSTIE}
{\INSTEG}
{\INSTBB}
{\INSTI}
{\INSTHE}
{\INSTDI}
{\INSTBA}
{\INSTJB}
{\INSTHH}
{\INSTPS}
{\INSTED}
{\INSTBL}
{\INSTII}
{\INSTBF}
{\INSTB}
{\INSTGB}
{\INSTJA}
{\INSTIC}
{\INSTHA}
{\INSTBE}
{\INSTEH}
{\INSTEA}
{\INSTIF}
{\INSTFJ}
{\INSTDH}
{\INSTGD}
{\INSTJP}
{\INSTCE}
{\INSTJG}
{\INSTDJ}
{\INSTJD}
{\INSTHJ}
{\INSTCG}
{\INSTH}
{\INSTHI}
{\INSTCF}
{\INSTHG}
{\INSTHC}
{\INSTFE}
{\INSTJC}
{\INSTGJ}
{\INSTDF}
{\INSTHD}
{\INSTJE}
{\INSTBD}
{\INSTIP}
{\INSTHF}
{\INSTGA}
{\INSTJH}
{\INSTIG}
{\INSTID}
{\INSTGC}
{\INSTVI}
{\INSTBC}
{\INSTFH}
{\INSTNB}
{\INSTJF}
{\INSTIA}
{\INSTCU}
{\INSTD}

\author{S.~Abubakar}
\affiliation{\Erciyes}

\author{M.~A.~Acero}
\affiliation{\Atlantico}

\author{B.~Acharya}
\affiliation{\Mississippi}

\author{P.~Adamson}
\affiliation{\FNAL}

\author{N.~Anfimov}
\affiliation{\JINR}

\author{A.~Antoshkin}
\affiliation{\JINR}

\author{E.~Arrieta-Diaz}
\affiliation{\Magdalena}

\author{L.~Asquith}
\affiliation{\Sussex}

\author{A.~Aurisano}
\affiliation{\Cincinnati}

\author{D.~Azevedo}
\affiliation{\UFG}

% ...

\collaboration{The NOvA Collaboration}
\noaffiliation

\author{K.\,Abe}\INSTBJ
\author{S.\,Abe}\INSTBJ
\author{H.\,Adhkary}\INSTWP
\author{R.\,Akutsu}\INSTCB
\author{H.\,Alarakia-Charles}\INSTEJ
\author{Y.I.\,Alj Hakim}\INSTFB
\author{S.\,Alonso Monsalve}\INSTEF
\author{L.\,Anthony}\INSTEI
\author{S.\,Aoki}\INSTCC
\author{K.A.\,Apte}\INSTEI
% ...

\collaboration{The T2K Collaboration}\noaffiliation

\maketitle

\section*{Abstract}
\vspace{-0.3cm}
The landmark discovery that neutrinos have mass and can change type (or ``flavor'') as they propagate---a process called neutrino oscillation~\cite{Super-Kamiokande:1998kpq,Super-Kamiokande:2002ujc,SNO:2001kpb,SNO:2002tuh,KamLAND:2002uet, DayaBay:2012fng}---has opened up a rich array of theoretical and experimental questions being actively pursued today. Neutrino oscillation remains the most powerful experimental tool for addressing many of these questions, including whether neutrinos violate charge-parity (CP) symmetry, which has possible connections to the unexplained preponderance of matter over antimatter in the universe~\cite{Fukugita:1986hr,Buchmuller:2005eh,pascoli2007connecting,branco2012leptonic,hagedorn2018cp}.  Oscillation measurements also probe the mass-squared differences between the different neutrino mass states ($\Delta m^2$), whether there are two light states and a heavier one (normal ordering) or vice versa (inverted ordering), and the structure of neutrino mass and flavor mixing~\cite{ParticleDataGroup:2022pth}.  Here, we carry out the first joint analysis of data sets from NOvA~\cite{NOvA:2021nfi} and T2K~\cite{T2K:2023smv}, the two currently operating long-baseline neutrino oscillation experiments (hundreds of kilometers of neutrino travel distance), taking advantage of our complementary experimental designs and setting new constraints on several neutrino sector parameters. This analysis provides new precision on the $\Delta m^2_{32}$ mass difference, finding $2.43^{+0.04}_{-0.03}\ \left(-2.48^{+0.03}_{-0.04}\right)\times 10^{-3}~\mathrm{eV}^2$ in the normal (inverted) ordering, as well as a $3\sigma$ interval on $\delta_{\rm CP}$ of $[-1.38\pi,\ 0.30\pi]$ $\left([-0.92\pi,\ -0.04\pi]\right)$ in the normal (inverted) ordering. The data show no strong preference for either mass ordering, but notably if inverted ordering were assumed true within the three-flavor mixing paradigm, then our results would provide evidence of CP symmetry violation in the lepton sector.  

\section*{Main}
The Standard Model of particle physics, extended to include neutrino mass, describes three flavor eigenstates of neutrinos ($\nu_e$, $\nu_\mu$, $\nu_\tau$) that are related to three mass eigenstates ($\nu_1$, $\nu_2$, $\nu_3$) by a $3\mathord{\times}3$ complex unitary mixing matrix $U_\mathrm{PMNS}$~\cite{Maki:1962mu,Pontecorvo:1967fh,Mohapatra_2007}.  This mixing, together with non-zero neutrino mass, allows for the phenomenon of neutrino oscillation whereby, during propagation, the flavor content of a neutrino evolves at a rate that depends on neutrino mass-squared splittings ($\Delta m^2_{ij} \equiv m_i^2 - m_j^2$) and the $U_\mathrm{PMNS}$ matrix elements.  In addition to these oscillation parameters, the rate depends on neutrino energy $E_\nu$ and neutrino propagation distance $L$ (``baseline'').  While experiments studying this process in recent decades have yielded great insights into the details of neutrino masses and mixings~\cite{ParticleDataGroup:2022pth}, many open questions remain.

The mixing matrix $U_\mathrm{PMNS}$ is typically parameterized in terms of three mixing angles ($\theta_{12}$, $\theta_{13}$, $\theta_{23}$) and at least one complex phase $\delta_{\rm CP}$~\cite{ParticleDataGroup:2022pth}.  It is unknown whether $\sin \delta_{\rm CP}$ is non-zero; if it is, neutrinos---and thus leptons---violate charge-parity (CP) symmetry and thereby provide a source of matter/antimatter asymmetry in nature~\cite{Mohapatra_2007}, which is of great interest given the connection between CP violation and the unexplained matter dominance in the universe~\cite{Fukugita:1986hr,Buchmuller:2005eh,pascoli2007connecting,branco2012leptonic,hagedorn2018cp}.  Separately, oscillation experiments have established that the mass-squared splitting $\Delta m^2_{32}$ is roughly thirty times larger in magnitude than $\Delta m^2_{21}$, but the sign of the former is currently unknown.  That is, $\nu_3$ may be heavier or lighter than the $\nu_1$/$\nu_2$ pair, with these two options termed respectively the normal ($\Delta m^2_{32}\mathord{>}0$) and inverted ($\Delta m^2_{32}\mathord{<}0$) mass orderings. 
Knowledge of the mass ordering can constrain experimental searches and theory development in a wide range of physics, including absolute neutrino mass measurements~\cite{formaggio2021direct}, neutrinoless double beta decay searches to investigate the nature of neutrino mass~\cite{dolinski2019neutrinoless}, models of supernova explosion and detection~\cite{hansen2020timing,horiuchi2018can}, and the cosmological evolution evidenced in cosmic microwave background and large scale structure measurements~\cite{lesgourgues2012neutrino}.
For the mixing angles, current data suggest $\theta_{23}$ is near $45^\circ$, a notable value hinting at a $\mu/\tau$ flavor symmetry~\cite{Mohapatra_2007}.  Improved precision on this and other mixing angles is essential in gaining a clearer view of flavor mixing and to probe the validity of the three-flavor paradigm.

Long-baseline accelerator neutrino oscillation experiments are well suited to address the above questions. In these, a high-intensity neutrino beam enriched in muon neutrinos ($\nu_\mu$) or muon antineutrinos ($\bar{\nu}_\mu$) is produced at a particle accelerator and directed through the Earth's crust towards a massive far detector located hundreds of kilometers away. Note that the word ``neutrino'' is used to mean both neutrino and antineutrino unless stated otherwise. The far detector measures the event rates of $\nu_\mu$ and $\nu_e$---the latter primarily from $\nu_\mu\mathord{\rightarrow}\nu_e$ oscillation---as a function of neutrino energy, from which the oscillation parameters above can be determined. These experiments employ near detectors, sited a short distance from the beam source, where oscillation effects are negligible and a very high neutrino event rate can be measured. The near detectors provide vital control measurements that significantly mitigate large systematic uncertainties in the initial neutrino flux, neutrino-on-nucleus interaction cross sections, and in some cases detector response ({\em e.g.}, energy reconstruction and event selection efficiencies).

Two such experiments are in operation today, T2K and NOvA. Each experiment uses a narrow-band off-axis beam~\cite{beavis1995long, Helmer:1994ac} whose peak energy is near the first oscillation maximum, $\sin^2\left(\frac{\Delta m_{32}^2 L}{4E}\right) \approx 1$, at the far detector. Note that natural units, where  $\hbar\,\mathord{=}\,c\,\mathord{=}\,1$, are used throughout. T2K uses an  $\sim$0.6~GeV neutrino beam from J-PARC in Tokai, Japan, and the 50-kt Super-Kamiokande water Cherenkov detector for its far detector located 295 km away~\cite{T2K:2011qtm}.  In the U.S., NOvA's $\sim$2~GeV beam is produced at Fermilab near Chicago, and the 14-kt tracking calorimeter far detector is located 810~km away in northern Minnesota~\cite{NOvA:2007rmc}.    Additional details on the designs of NOvA, T2K, and on long-baseline experiments generally can be found in the Methods and in Refs.~\cite{T2K:2011qtm, NOvA:2007rmc, DiLodovico:2023jgr}.

We report here a combined analysis of the data sets from T2K and NOvA previously analyzed independently in Refs.~\cite{T2K:2023smv,NOvA:2021nfi}. This combination takes advantage of significant complementarity in the two experiments’ sensitivities to the oscillation parameters. In particular, the $\nu_\mu\mathord{\rightarrow}\nu_e$ oscillation probability is a function of (among other things) both $\delta_{\rm CP}$ and the neutrino mass ordering, and these two effects must be teased apart.

Figure 1 illustrates the complementarity between the experiments in a simplified case. Sets of oval curves indicate the energy-integrated total $\nu_e$ and $\bar{\nu}_e$ event counts  expected in the far detectors under various mass ordering and $\delta_{\rm CP}$ scenarios, with other oscillation parameters held fixed. The measured event counts in NOvA and T2K are  shown as black points with error bars.

As shown in the top panel, there is stronger separation between the mass ordering ovals for NOvA, due to higher beam energies, but since NOvA’s data lie near the overlap of the ellipses, there can be ambiguity as to which ordering is the correct one and (in a correlated way) which values of $\delta_{\rm CP}$ are preferred. In contrast, T2K has less sensitivity to the mass ordering, but points with similar values of $\delta_{\rm CP}$ in each hierarchy sit close to one another, and the data lie closest to $\delta_{\rm CP}=-\frac{\pi}{2}$ regardless of mass ordering. Combining these data sets can provide simultaneous mass ordering and $\delta_{\rm CP}$ information with substantially less ambiguity, maximizing the utility of current data and informing data-taking strategies for current and future experiments.

\begin{figure}
    \centering
    \includegraphics[width=.5\textwidth]{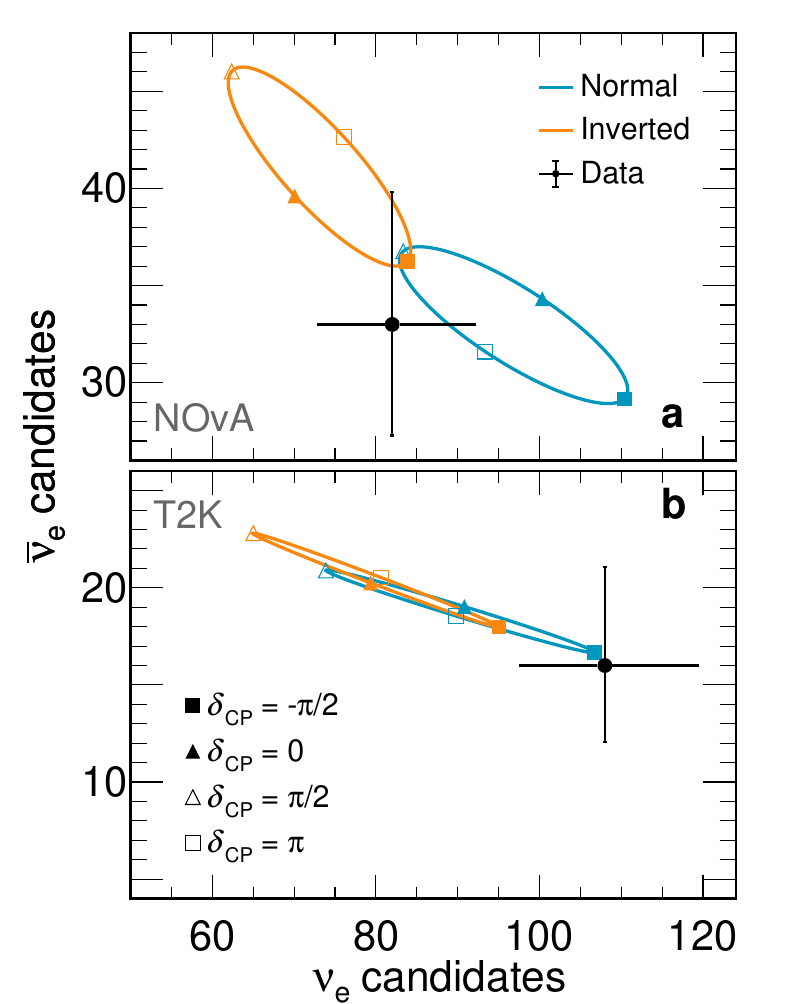}
    \caption{\textbf{The impact of mass ordering and $\delta_{\rm CP}$ on event rates.} A ``bi-event'' plot that illustrates experimental sensitivity to neutrino mass ordering and $\delta_{\rm CP}$, with panels representing the NOvA (a) and T2K (b) cases.   Black points with $1\sigma$ Poisson statistical error bars show the total number of $\nu_e$ and $\bar{\nu}_e$ candidates selected in the far detectors.  The oval parametric curves trace out predicted numbers of events under the normal (blue) or inverted (orange) mass ordering assumption as the parameter $\delta_{\rm CP}$ varies from $-\pi$ to $\pi$.  Four specific $\delta_{\rm CP}$ values are labeled for reference.  All other oscillation parameters are kept fixed in this graphic, set to their most probable values from the joint analysis (Extended Data Table~\ref{tab:hpds}.)}
    \label{fig:bievent}
\end{figure}

This discussion points to a more general observation that the oscillation parameters of interest represent a highly correlated multidimensional space. The analysis reported here calculates a joint Bayesian posterior, using the experiments' likelihoods defined over the full parameter space.  Additionally, we utilize the full suite of analysis tools from both experiments: detector response models, neutrino energy estimators, near detector measurements, and systematic uncertainties, all within a unified framework for statistical inference. This level of integration is a first for accelerator neutrino experiments.

The posterior calculation is based on detailed parameterized models of the neutrino flux, cross sections, and detectors that predict the binned spectra of neutrino events in each of our selected samples as a function of the oscillation parameters and a large number of nuisance parameters mostly related to systematic uncertainties in the models. A likelihood is constructed from Poisson probability terms describing the compatibility between the prediction and the observed data in bins of relevant variables. Prior probabilities are set on all parameters as detailed in the Methods.

Both T2K and NOvA have software that explores the posterior using Markov chain Monte Carlo (MCMC) methods~\cite{Metropolis:1953am,Hastings:1970aa} (ARIA for NOvA~{\cite{NOvA:2023iam}} and MaCh3 for T2K~\cite{the_mach3_collaboration_2024}). By containerizing~\cite{kurtzer2017singularity} the likelihood and prior portions of the code, we are able to construct and analyze the joint posterior using either of the original MCMC frameworks, despite the very different software environments involved. For each fitting framework, ARIA or MaCh3, the fitter's native likelihood and priors are calculated directly while the other experiment's likelihood and priors are accessed via the software container.  In this way, either framework can be used, providing valuable redundancy and thus cross checks of all statistical inferences.

While a single set of oscillation parameters naturally applies to both experiments in the joint posterior, the treatment of the many nuisance parameters related to systematic uncertainties is more subtle. Both measurements of the oscillation parameters currently have statistical uncertainties larger than the systematic uncertainties, but the latter are not negligible. We thoroughly surveyed the flux, cross-section, and detector models and their systematic uncertainties to determine whether correlations between the experiments affect the analysis at a significant level. Our conclusions from this effort are summarized in the following paragraphs.

Both T2K and NOvA use beams produced by directing accelerated protons onto graphite targets. The hadrons are charge selected with magnetic horns: positively charged hadrons decay to produce neutrinos and negatively charged to produce antineutrinos. Many uncertainties on these beam fluxes stem from processes unrelated between the two experiments, {\em e.g.}, alignment of beam components. Yet, uncertainties on the rate of hadron production in the graphite targets are significant, and the underlying physics is the same. However, the two experiments use proton beams of different energies (30~GeV for T2K, 120~GeV for NOvA), and the external data sets used to tune the hadron production models of both experiments are different~\cite{NA49:2006oyk,NA61SHINE:2016nlf,NA61SHINE:2015bad}.  Additionally, the ultimate impact of flux uncertainties on far detector predictions in NOvA is much smaller than other uncertainties. We therefore conclude that at current experimental exposures the two experiments' flux uncertainties need not be correlated.

Given the different detector technologies involved, most detector-related uncertainties are independent between the experiments.  
Furthermore, the very different energy estimation techniques used, combined with model tuning and uncertainty estimation using in-situ calibration samples in each experiment, including for the lepton and neutron energy scales, leads to independence between the two detector uncertainty models. 
We conclude that there are no significant correlations in the detector models.

For neutrino-on-nucleus cross sections, the underlying physics is the same, and in many cases the same external data sets are used by both experiments to tune and set prior uncertainties on model parameters. Thus, cross-section model correlations are expected. However, in the specific case of NOvA and T2K, the description of this physics differs in significant ways.  The simulation packages differ~\cite{Hayato:2021heg,Andreopoulos:2009rq}, the physical models implemented in them differ in many places, the parameterizations differ almost entirely, and customized tunings are necessary and applied given the experiments’ specific energies, detector technologies, and approaches to systematic uncertainty mitigation. 

Proper correlations between experiments could be implemented by starting from a common cross-section model spanning different energy ranges and able to describe both the leptonic and hadronic parts of the final state. Such a joint description is not yet mature and is one of the focuses of the community in the years to come~\cite{Balantekin:2022jrq}. Given the differences in the models, a direct mapping of their parameters was deemed not practical at this time. Instead, we studied whether neglecting these correlations could appreciably affect our measurements of the oscillation parameters. The studies are limited to our current experimental exposures and models and would need reevaluation if applied to any other context.

First, we assessed whether correlations between single systematic parameters in our models could have a significant impact on our results. For each of $\Delta m^2_{32}$, $\theta_{23}$ and $\delta_{\rm CP}$, we identified the systematic parameter on each experiment with the largest impact on that oscillation parameter's measurement.  Then, regardless of whether those two systematic parameters made physical sense to correlate, we performed fits to simulated pseudo-data with the parameters fully correlated, uncorrelated, and fully anticorrelated. Details of these studies including how we identified the most impactful parameters are shown in~the~Methods. In summary, we saw no case where the choice of correlation of individual systematic parameters  significantly affected the oscillation parameter measurements.

Checking individual parameters does not rule out effects from a mix of systematic parameter variations that combine to produce a net effect that is larger and possibly more degenerate with oscillation effects, representing a potential worst-case scenario for the analyses. Rather than seeking such a set of variations, we directly identified, or in some cases constructed, single systematic parameters for each experiment that have effects similar to each oscillation parameter of interest. We then adjusted the size of the priors on these ``nightmare'' parameters such that their impact on the measurements is comparable to that of statistical errors and therefore larger than the net effect of all our regular systematic parameters combined. These nightmare parameters were added to our nominal uncertainty models to create ``augmented models'' allowing us to study a case where systematic effects are comparable to statistical uncertainty. Next, we constructed simulated pseudo-data sets with the nightmare parameters increased in both experiments by one standard deviation above their prior central values.  These simulated pseudo-data were then fit three times using the augmented model: once with the experiments' nightmare parameters fully correlated (matching the pseudo-data), once fully anticorrelated, and finally uncorrelated. We find that the oscillation parameter constraints extracted in the fully correlated and uncorrelated cases have negligible differences. However, the incorrect anticorrelated case yields a large bias. We expect that with even larger systematic uncertainties, differences between the correlated and uncorrelated case would eventually become relevant. However, this study indicates that we are not in such a regime with the current exposures and systematic uncertainties. See Methods for further results. 

Given that no significant biases are seen from neglecting correlations between actual systematic parameters, and the only bias seen with the nightmare parameters comes not from neglecting a correlation but by adding an incorrect one, we choose in most cases to neglect the correlations between the two experiments' systematic uncertainties. The one exception relates to the $\approx 2\%$ normalization uncertainties on all $\nu_{e}$ and $\bar{\nu}_{e}$ events described in~\cite{day2012differences}. In this case, the uncertainties are implemented identically by T2K and NOvA, and we have correlated them.

We also perform studies wherein the joint fit is tested against pseudo-data constructed with a set of discrete model variations not directly accessible using the experiments' nominal uncertainty models. This procedure was used in the earlier independent T2K analysis~\cite{T2K:2023smv}, and we include in the present analysis those model variations seen as most impactful previously. Similarly, we studied a secondary set of variations based on extrapolating each experiment's cross-section model to the other experiment's context. Predefined thresholds were used to establish that no substantive changes in the oscillation parameters' central values or interval widths were seen under these tests, as described in the Methods.    For all tested alternative models, all observed changes in credible intervals were within thresholds.  See Methods for further details.  Each experiment naturally continues to investigate improvements in their cross-section models, and the studies described here would warrant repeating for larger data exposures and/or updated theoretical understanding. Continued theoretical and experimental effort in this direction is important.

With the joint likelihood and systematic uncertainty model defined, we use our fitting frameworks to analyze the combined data sets of Refs.~\cite{T2K:2023smv,NOvA:2021nfi}, finding consistent results between the two frameworks. Unless stated otherwise, we report results using an external constraint on $\theta_{13}$ (named the ``reactor constraint'' below) and external constraints on $\Delta m^2_{21}$ and $\theta_{12}$.  The values used for these constraints correspond to the 2020 Particle Data Group summary values~\cite{ParticleDataGroup:2020ssz} and are given in Methods.

We tested the goodness-of-fit (see Methods) of our model to data using the $p$-value method~\cite{gellmanpospred}, both overall and for each individual sample in the far detectors. All the $p$-values are within an acceptable range ($>$0.05 after the look-elsewhere-effect adjustment described in the Methods). The overall $p$-value to describe all NOvA and T2K samples is 0.75 (0.40) for full spectral (rate-only) analysis, marginalized over both mass orderings. Similar results were obtained without the reactor constraint and in each mass ordering. Thus the joint oscillation model simultaneously fits T2K and NOvA data well. The $p$-values are also consistent with those of previous T2K-only and NOvA-only analyses.

We produce parameter estimations via highest-posterior-density credible intervals and perform discrete hypothesis tests using the Bayes factor formalism. Conclusions related to CP conservation/violation, $\Delta m^2_{32}$, $\sin^2\theta_{23}$ and mass ordering have been tested to be robust under the alternative model variations described previously. For the measured oscillation parameters we report $1\sigma$ ($68.27\%$) credible intervals unless noted.

\begin{figure}
    \centering
    \includegraphics[width=.5\textwidth]{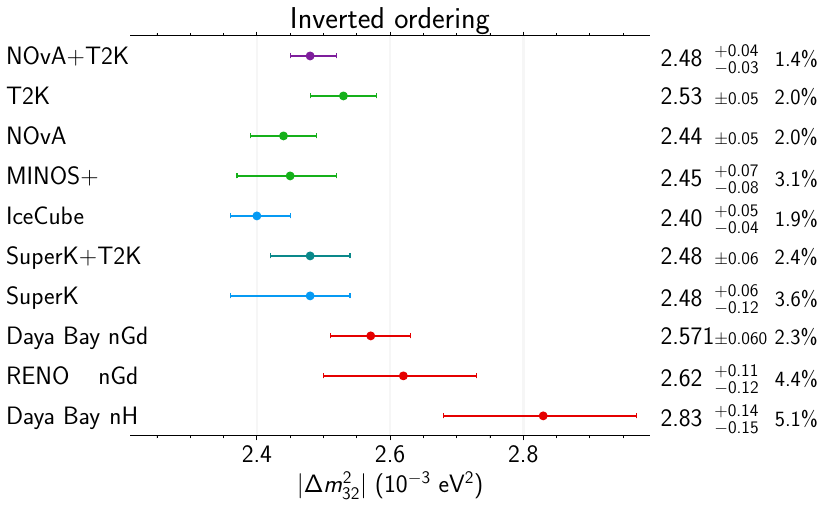}
    \caption{\textbf{Experimental measurements of $\vert \Delta m^2_{32}\vert$.} The measurements assume the inverted ordering preferred by this analysis. Sources for the results from top to bottom starting with the second line are as follows:\ \cite{T2K:2023smv, NOvA:2021nfi, MINOS:2020llm, IceCubeCollaboration:2024zec,abe2024first, Super-Kamiokande:2023ahc, DayaBay:2022orm, RENO:2024msr, DayaBay:2024hrv}. The normal ordering case is available in Extended Data Fig.~\ref{fig:global}.}
    \label{fig:global_ih}
\end{figure}

We find $\sin^2 \theta_{23} = 0.56_{-0.05}^{+0.03}$ without any assumptions on the ordering of the neutrino masses. The fit weakly prefers the upper octant of $\theta_{23}$ ($\sin^2\theta_{23}>0.5$) over the lower octant with a Bayes factor of 3.5. Removing the reactor constraint gives no statistically significant preference for either octant (Bayes factor 1.2 for lower octant vs upper octant).  We also find $\Delta m_{32}^2 = 2.43_{-0.03}^{+0.04}~\left(-2.48_{-0.04}^{+0.03}\right) \times 10^{-3}$~eV$^2$ assuming the normal (inverted) ordering. This is currently the smallest experimental uncertainty  on $\vert \Delta m_{32}^2 \vert$ (Fig.~\ref{fig:global_ih}). This conclusion also applies when the reactor constraint is replaced by a flat prior.

There is no statistically significant preference obtained for either of the mass orderings with a Bayes factor of 1.3 (2.5) in favor of the inverted ordering with (without) reactor $\theta_{13}$ constraint. While the two experiments individually prefer the normal ordering, the values of other oscillation parameters are more consistent in the inverted ordering, leading to a different ordering preference in the joint fit, though still not statistically significant.  The effect on mass ordering preference when additionally incorporating reactor $\Delta m_{32}^2$ measurements is discussed in the Methods.

\begin{figure*}
    \centering
    \includegraphics[width=\textwidth]{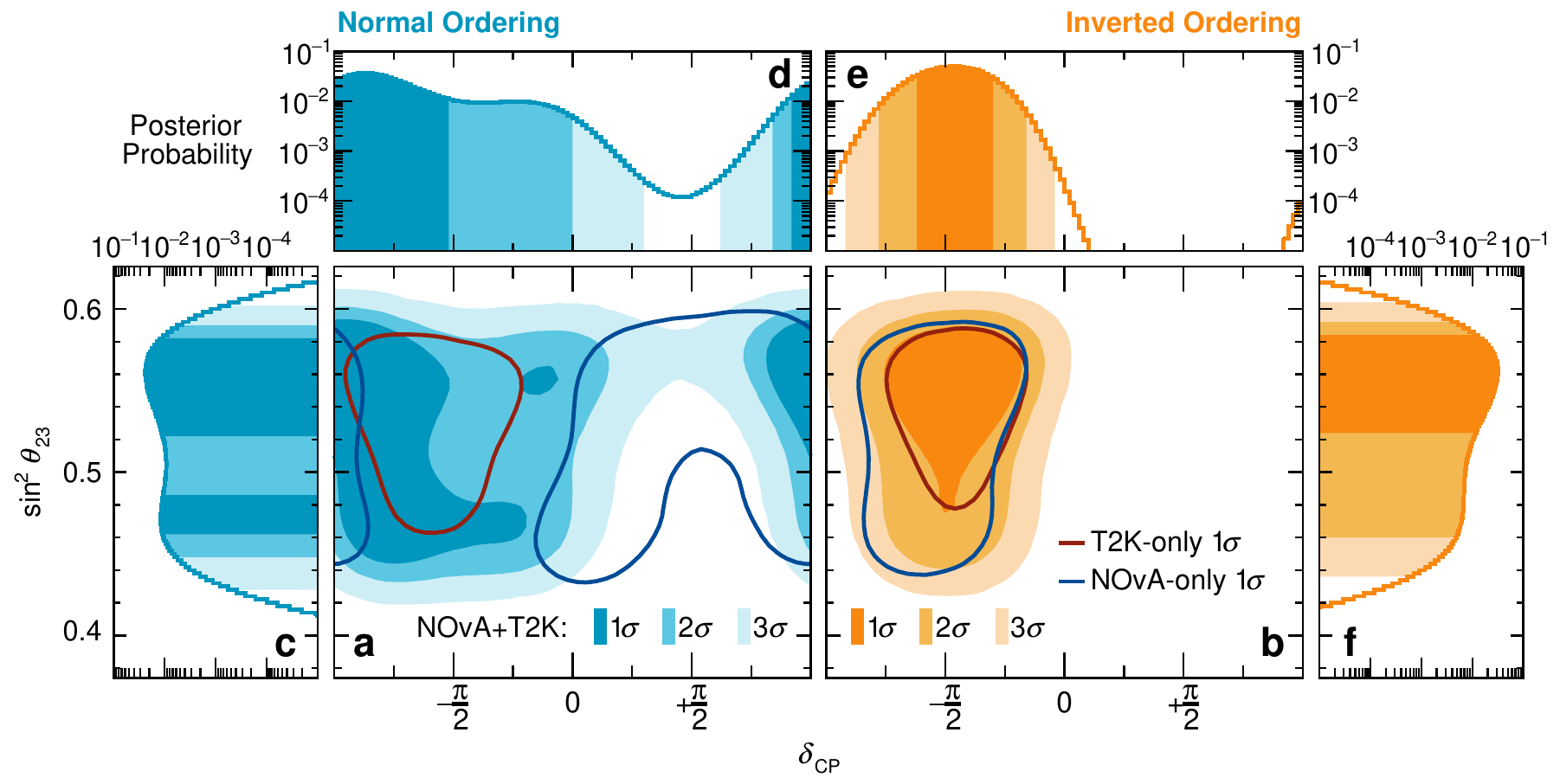}
    \caption{\textbf{Constraints on $\sin^2\theta_{23}$ and $\delta_{\rm{CP}}$.} Marginalized posterior probabilities and 1D or 2D Bayesian credible regions of $\sin^2 \theta_{23}$ and $\delta_{\rm CP}$ in the case of the normal (blue, left side) and inverted (orange, right side) neutrino mass ordering with the reactor constraint applied. Shaded areas correspond to $1 \sigma$, $2 \sigma$, and $3 \sigma$ credible regions. 2D panels of $\sin^2 \theta_{23}$ vs $\delta_{\rm CP}$ (a, b) are overlaid with $1 \sigma$ credible regions from the T2K-only (dark red) and NOvA-only (dark blue) data fits assuming normal (a) and inverted ordering (b). 1D panels show the posterior probabilities of $\sin^2 \theta_{23}$ (c) and $\delta_{\rm CP}$ (d) in the normal ordering, and $\delta_{\rm CP}$ (e) and $\sin^2 \theta_{23}$ (f) in the inverted ordering.}
    \label{fig:dcp_th23}
\end{figure*}

\begin{figure*}
    \centering
    \includegraphics[width=\textwidth]{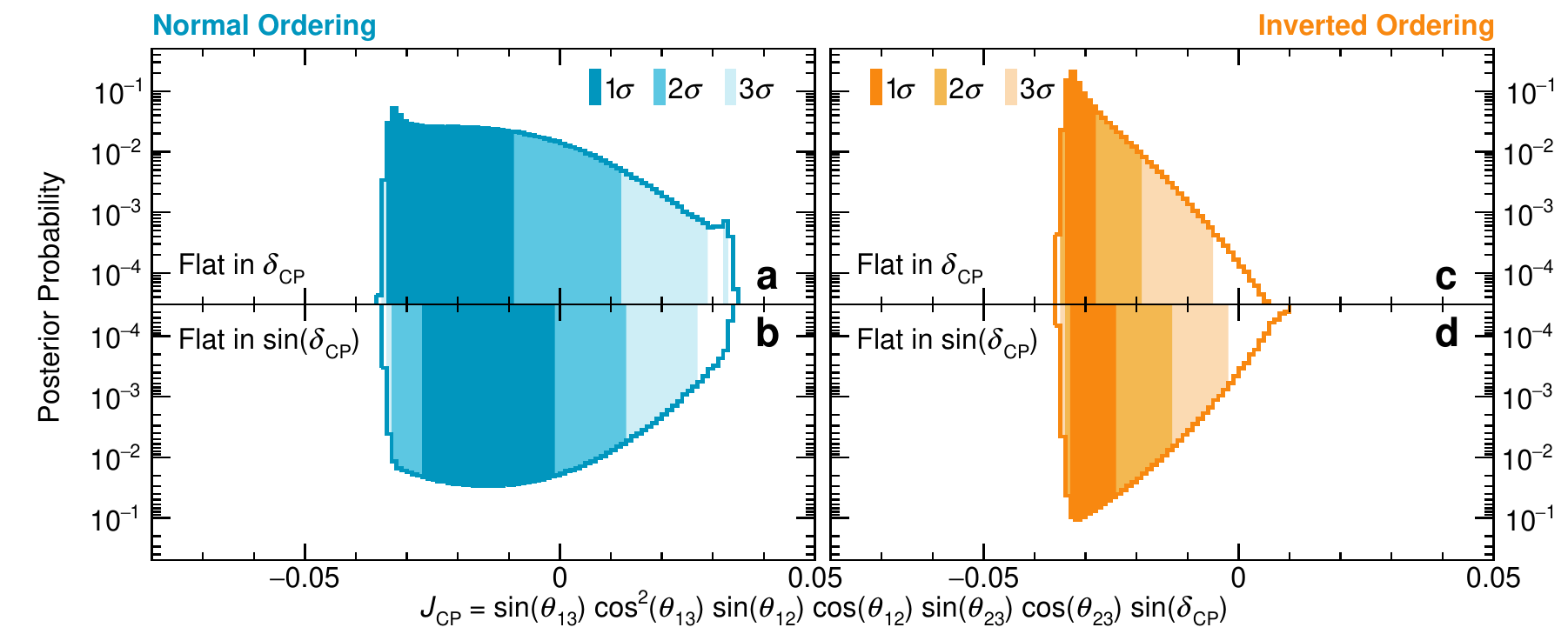}
    \caption{\textbf{Constraints on the Jarlskog invariant.} Marginalized posterior probabilities of the Jarlskog invariant, $J_{\rm CP}$, in the case of the normal (blue; a, b) and inverted (orange; c, d) neutrino mass ordering with the reactor constraint applied. The posterior distributions use prior distributions either flat in $\delta_{\rm CP}$ (a, c) or $\sin\delta_{\rm CP}$ (b, d). Shaded areas show the $1 \sigma$, $2 \sigma$, and $3 \sigma$ Bayesian credible intervals.}
    \label{fig:jarlskog}
\end{figure*}

With no assumption on the true mass ordering, we find the $1\sigma$ credible interval on $\delta_{\rm CP}$ to contain $[-0.81\pi,\ -0.26\pi]$ with the highest posterior probability value being $-0.47\pi$. We also find that values of $\delta_{\rm CP}$ around $+\pi/2$, an extremum of $\sin \delta_{\rm CP}$, are outside our $3\sigma$ ($99.73\%$) credible intervals, which also holds for either mass ordering separately. Figure~\ref{fig:dcp_th23} shows the joint fit result compared to NOvA and T2K's individual measurements in the $\sin^2 \theta_{23}$-$\delta_{\rm CP}$ plane as well as 1D uniformly binned posterior probability distributions for both mass ordering cases. Assuming the normal ordering, the joint analysis allows a wide range of $\delta_{\rm CP}$ values giving a $3\sigma$ credible interval of $\delta_{\rm CP} \in [-1.38\pi,\ 0.30\pi]$. In the case of the inverted ordering $\delta_{\rm CP} \in [-0.92\pi,\ -0.04\pi]$, excluding $56\%$ of the parameter space; the CP-conserving values of $\delta_{\rm CP} = 0$ and $\pi$ are outside the $3\sigma$ credible interval. A consistent picture is seen when analyzing the Jarlskog invariant, $J_{\rm CP}$~\cite{jarlskog}, which is a parametrization-independent measure of CP-violation. The CP-conserving value of $J_{\rm CP}=0$ falls outside the $3\sigma$ credible interval for the inverted ordering, and the above statements are true whether the prior used is uniform in $\delta_{\rm CP}$ or $\sin\delta_{\rm CP}$ (Fig.~\ref{fig:jarlskog}). This analysis therefore provides evidence for CP violation in the lepton sector if the inverted ordering is assumed to be true. However, we do not currently see a significant preference for either mass ordering. Future mass ordering measurements will therefore influence the interpretation of these results.  See Methods for additional data projections and comparisons.

\clearpage
\section*{Methods}\label{sec:meth}
\renewcommand{\figurename}{Extended Data Fig.}
\renewcommand{\tablename}{Extended Data Table}
\renewcommand{\thefigure}{\arabic{figure}}
\setcounter{figure}{0}
\setcounter{table}{0}

\subsection*{The NOvA Experiment}\label{sec:meth:NOvAExperiment}
The NOvA experiment measures neutrino oscillations using two detectors of functionally identical construction located along the NuMI neutrino beam~\cite{Adamson:2015dkw} produced at Fermi National Accelerator Laboratory (Fermilab). 

The smaller 0.3-kt near detector is located on the Fermilab campus 1~km downstream from the neutrino production target, while the 14-kt far detector is located 810~km away in northern Minnesota.  The detectors themselves are highly segmented tracking calorimeters consisting of long PVC cells filled with a mineral-oil-based liquid scintillator.  Each cell measures 6.6~cm $\times$ 3.9~cm in cross section, runs the full height or width of the detector (15.5~m for the far detector and 3.9~m for the near detector), and is instrumented with a wavelength-shifting fiber and avalanche photodiode to detect the scintillation light produced when charged particles pass through the cell.  The cells are arranged in a series of layers, each with either horizontal or vertical orientation, with the direction alternating between layers to provide 3D event reconstruction.  This segmented design offers the excellent muon and electron classification needed for tagging the incoming neutrino flavor.  In particular, electromagnetic showers at typical NOvA energies are much larger than the detector cell widths and thus are well-imaged and distinct from many potential backgrounds.  NOvA's detectors are centered 14.6~mrad off the NuMI beam's central axis, yielding a narrow-band neutrino beam peaked at 1.8~GeV.
 
As is typical for particle physics experiments, NOvA makes use of detailed simulations of beam production, neutrino interaction physics, and detector response as part of the analysis.  Given the matching near and far detectors, NOvA forms its oscillation-dependent predictions of the far detector event rates directly from data using the millions of neutrino interactions recorded in the near detector.  This near-to-far ``extrapolation'' process is carried out as a function of multiple kinematic and event classification variables.  Uncertainties from the simulations have greatly reduced impact as they enter the oscillation fit only to the extent that they affect the mapping between expected near and far event rates, not the individual detectors' event rates themselves.  Uncertainties on the simulations are taken as the a priori uncertainties from, for instance, the external model constraints or other external data and are supplemented by additional model uncertainties where a priori coverage was deemed unsatisfactory.
 
Far detector data are fitted to the corresponding far detector predictions to extract oscillation parameter constraints.  These data are separated by beam mode ({\em i.e.}, neutrino- or antineutrino-dominated running) and further into $\nu_\mu/\bar{\nu}_\mu$ charged current and $\nu_e/\bar{\nu}_e$ charged current candidate samples using a convolutional neural network~\cite{Aurisano:2016jvx} whose inputs are the calibrated event images recorded by the detector cells.  Subsequent reconstruction of tracks and showers within each event provide kinematic information such as estimated neutrino energy.  Far detector $\nu_\mu/\bar{\nu}_\mu$ samples are analyzed in bins of neutrino energy and hadronic energy fraction.  The $\nu_e/\bar{\nu}_e$ samples are analyzed in bins related to event containment, event classification score, and neutrino energy.  More details on the analysis techniques, simulation packages, systematic uncertainties, and the overall NOvA experimental design can be found in Ref.~\cite{NOvA:2021nfi} and the references therein.

\subsection*{The T2K Experiment}\label{sec:meth:T2KExperiment}
The T2K experiment is composed of the J-PARC neutrino beam, a near site with multiple detectors, and the water Cherenkov detector Super-Kamiokande (SK) as the far detector. Full details of the experiment can be found in Ref.~\cite{T2K:2011qtm}.

The primary detector at the near site, 280~m from the target, is a magnetized off-axis (centered at 43.6~mrad) tracking detector called ND280. While taking the data used in this analysis, ND280 consisted of a $\pi^0$ detector followed by a tracker consisting of three time-projection chambers (TPCs) interleaved with two hydrocarbon fine-grained detectors (FGD1 and FGD2), all surrounded by an electromagnetic calorimeter. The stability and direction of the neutrino beam are monitored using the on-axis near detector INGRID.

SK is situated 295 km downstream of the neutrino production target, 43.6 mrad off-axis, and contains 50~kt of water. An inner detector (ID) using 11,129 inward-facing 20-inch photomultiplier tubes (PMTs) detects Cherenkov radiation from charged particles traversing the detector. An optically separated outer detector uses 1885 outward-facing 8-inch PMTs to reject interactions originating outside the ID volume. SK is able to discriminate between electrons and muons by their Cherenkov ring profiles.

T2K uses a ``forward-fitting'' analysis strategy. First, a model that predicts the event spectra at the near and far detectors is defined and tuned to external experimental data. The predictions are generated by simulating the neutrino flux and cross section as well as the detector response. The model, with variable parameters, is fit to the ND280 data to obtain tuned values of the parameters with uncertainties. The constrained model resulting from this near detector fit is then used to make SK predictions, which are fit to the SK data to extract oscillation parameters. Complete details for this analysis, including model details, are in Ref.~\cite{T2K:2023smv}.

T2K splits data at the near and far detectors into mutually exclusive samples defined by particle identification in each beam mode. At ND280, events are categorized into 18 samples, nine samples in each of FGD1 and FGD2. In neutrino-mode, data with one negatively charged muon is split into three samples in each FGD corresponding to the number of pions (0, 1, or $>$1). In antineutrino mode data are first split by whether a negatively or positively charged muon is present, and then divided by number of pions as in the neutrino-mode data, forming six samples in each FGD. For all samples, the data are fit in a 2D space of the muon momentum and the angle between the muon and the average beam direction. The exclusive samples allow the near detector fit to better constrain parameters related to different neutrino-nucleus interaction modes. At SK, the data are divided into three samples in neutrino-mode: 1-ring muon-like, 1-ring electron-like, and 1-ring electron-like with one decay electron; in antineutrino mode, only the 1-ring muon-like and 1-ring electron-like samples are used. The data are binned in reconstructed neutrino energy. All electron-like samples are additionally binned in a second dimension, the angle between the reconstructed electron direction and the beam direction. 

Detector systematic uncertainties are evaluated using a variety of sideband samples and calibrations, covering effects such as particle identification, particle momentum reconstruction, secondary particle interactions, and fiducial volume effects.

\subsection*{Correlations in Flux Modeling}
The modeling of the neutrino flux depends on many details relating to the incident proton beam, the hadron production target, and the magnetic focusing horns. As these details are specific to each experiment, flux systematic uncertainties due to magnetic field variations, component alignment, and other beamline properties are uncorrelated between the experiments. 

The only possible correlation identified was the pion and kaon production models and the use of hadron interaction experiments to tune them~\cite{T2K:2012bge, MINERvA:2016iqn}. In the case of NOvA, the primary data are from the NA49 experiment~\cite{NA49:2006oyk}, which collected thin-target  (slices of the target material) data at 158~GeV/c, which is then scaled to the NuMI beam energy. The NA61/SHINE experiment, which collected data for T2K, uses some of the same detectors and the same beamline as NA49. NA61/SHINE~\cite{NA61SHINE:2016nlf,NA61SHINE:2015bad} collected both thin-target and replica-target (a full-sized target) data for T2K at 31~GeV/c, the J-PARC beam momentum. Checking the consistency of the NA49 and NA61/SHINE data used is difficult since the data are collected at different beam energies. 

The NOvA experiment primarily uses thin-target NA49 hadron production data to tune the particle multiplicities, reweighting interactions and particle propagation inside the target and other beamline materials. On the other hand, T2K uses thin and replica-target data from NA61/SHINE to reweight the multiplicities of particles exiting the target. Given these differences in data collection and tuning methodology, and given that flux uncertainties have greatly suppressed influence after ND data constraints are considered, there is no expectation of significant correlations between flux systematic parameters for NOvA and T2K in the joint fit.

\subsection*{Correlations in Detector Modeling}
The experiments use different detector technologies as well as strategies for forming data samples, which removes most opportunities for correlation. However, the modeling of particle propagation through the detectors derives from the same underlying physics. This propagation is called secondary interaction (SI), and the case of pion SI is noteworthy, as this process is expected to occur in both experiments and for T2K is an important effect. T2K's selection uses exclusive samples where a change in reconstructed pion multiplicity can cause migration between samples. NOvA, on the other hand, uses inclusive selections, and pion SI has minimal effect on NOvA's calorimetric energy estimation. Thus, we do not expect significant correlations due to pion SI.

\subsection*{Tests of Individual Parameter Correlations}\label{sec:meth:corranticorr}
Neutrino-on-nucleus scattering plays a central role in both experiments, but the modeling of this physics has significant differences between the two individual analyses. These differences, together with the presence of different nuclear targets, neutrino energies, and near detector strategies, means that direct estimation of systematic uncertainty correlations in the neutrino scattering models is highly non-trivial. As part of this analysis, we tested how significant inter-experimental systematic uncertainty correlations could be, starting by identifying T2K's and NOvA's most impactful systematic uncertainties and exploring correlations between them.  

To determine an impactful systematic parameter, we carry out a fit to pseudo-data generated with all parameters at their prior values from our nominal model. Then for each parameter in turn, we reweight all steps from the obtained MCMC chain to have a tight (``shrunk'') prior for that parameter around a different value (``pulled'') to that used to generate the pseudo-data and study the change in the extracted oscillation parameter intervals. This procedure mocks up the result of an external experiment providing a strong constraint on each systematic parameter at a different value to that preferred by simulated pseudo-data. This ``shrink and pull'' study allows for assessing the single-parameter impact on the systematic uncertainty and the estimated credible intervals of the individual neutrino oscillation parameters' measurement.

First, we identify both NOvA's and T2K's systematic parameters with the largest impact on $\delta_{\rm CP}$, $\sin^2 \theta_{23}$ and $\Delta m_{32}^2$ in the joint fit. 

For both experiments the largest change in $\delta_{\rm CP}$ credible interval comes from uncertainties on $\nu_e$ and $\bar{\nu}_e$ normalizations. As discussed, these uncertainties are implemented identically in both experiments, and we have correlated them in the joint analysis. No additional interaction uncertainties in our models have any significant impact on the resulting credible intervals of $\delta_{\rm CP}$.

For $\sin^2 \theta_{23}$, all the individual interaction systematic parameters have very small effects, changing the width of the $1\sigma$ interval by less than 2\% when shrunk by 50\% and pulled $1\sigma$ away from the nominal value. The largest change in credible interval comes from the uncertainty on neutron visible energy for NOvA and 2p2h C/O cross-section scale for T2K (2p2h C/O cross-section scale allows the 2p2h cross-section on carbon to differ from that for oxygen). For $\Delta m_{32}^2$, all the individual interaction parameters have a negligible effect on the resulting $\Delta m_{32}^2$ credible intervals. Hence, we widened the list of considered parameters and identified NOvA's calorimetric energy scale uncertainty and T2K's SK energy scale uncertainty as the most impactful for $\Delta m_{32}^2$.

Second, despite there being no a priori reason to expect correlations between these specific parameters, we test whether or not correlating the most impactful T2K parameter with the most impactful NOvA parameter modifies oscillation parameter constraints in the joint fit in a significant way. We simulate pseudo-data to which we perform a joint fit while treating the T2K and NOvA parameters described above as either uncorrelated, fully correlated, or fully anticorrelated. We repeat the study for each pair of T2K's and NOvA's most impactful parameters with respect to $\delta_{\rm CP}$, $\sin^2 \theta_{23}$, and $\Delta m_{32}^2$. In the case of $\Delta m_{32}^2$, we further inflate the original SK energy scale uncertainty from 2\% to 7\% to amplify the effect. Finally, we check the extracted $1\sigma$ and $2\sigma$ credible regions for any substantial differences between the three correlation configurations. These tests are repeated for three sets of pseudo-data generated with oscillation parameter values that are T2K-like, NOvA-like, and NuFit-like~\cite{Esteban:2020cvm}, which are chosen to be close to recent data results from the respective collaborations and are given in Extended Data Table \ref{tab:asimovpoints}.

\begin{table}
\caption{\textbf{Default oscillation parameters for simulation.} Sets of oscillation parameter values used to generate pseudo-data. For all sets, $\sin^{2}\theta_{13}$ is $2.18\times 10^{-2}$, $\Delta m^{2}_{21}$ is $7.53\times 10^{-5}$~eV$^{2}$, and $\sin^{2}\theta_{12}$ is 0.307.}\label{tab:asimovpoints}
 \vspace{1ex}
 {
     \begin{tabular}{lccc}
     \hline
     & T2K-like & NOvA-like & NuFit-like \\
     \hline

          $\Delta$\textit{m}$^{2}_{32}$ (eV$^2$) & $2.51 \times 10^{-3}$  & $2.41 \times 10^{-3}$  & $-2.45 \times 10^{-3}$  \\
          $sin^{2}\theta_{23}$ & 0.528 & 0.570 & 0.550  \\ 
          $\delta_{\rm CP}$ & $-0.51\pi$ & $0.83\pi$ & $-0.50\pi$ \\
          \hline
     \end{tabular}}
\setlength{\baselineskip}{2\baselineskip}

\end{table}

\begin{figure*}
    \centering
    \includegraphics[width=\textwidth]{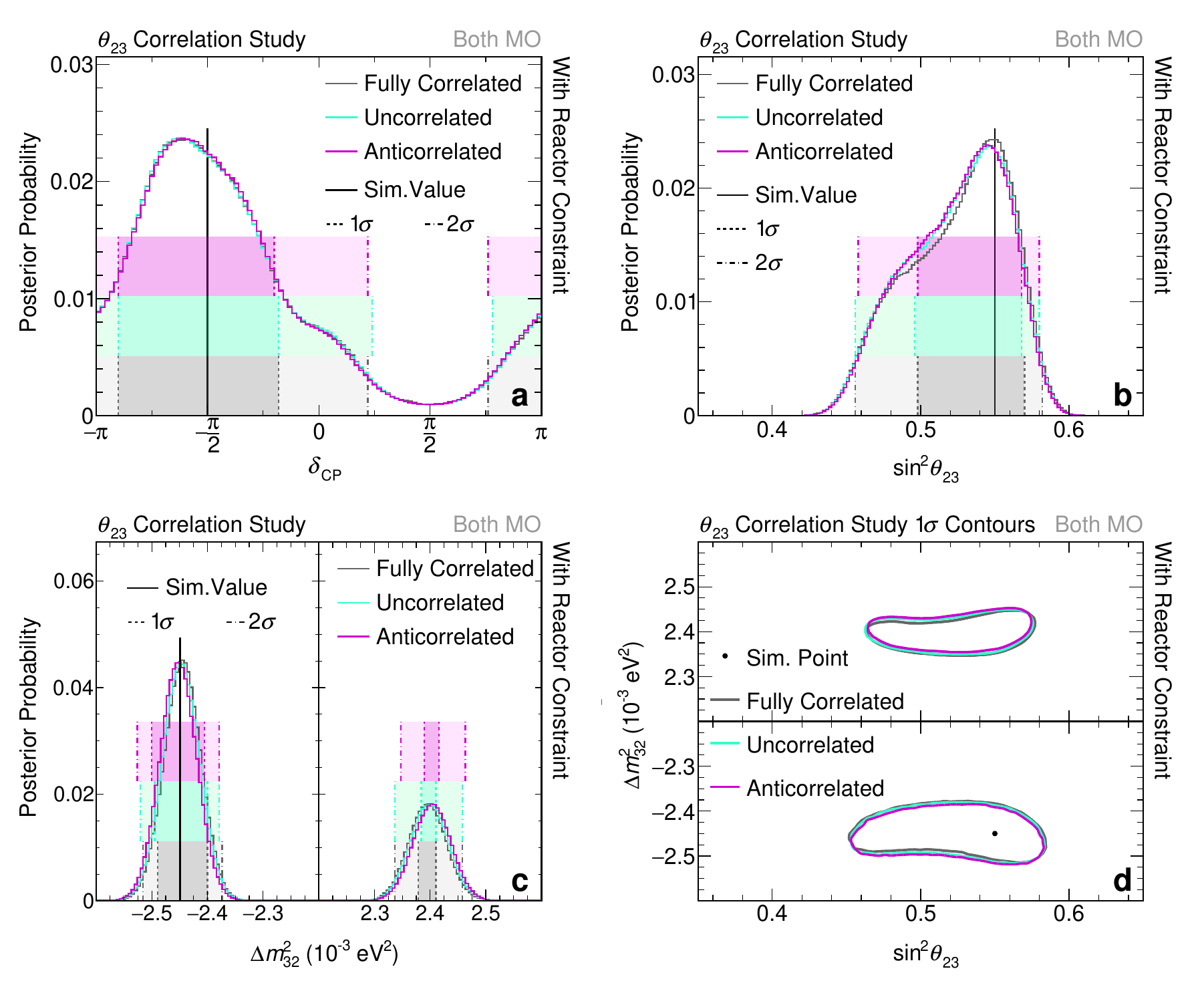}
    \caption{\textbf{Correlation study comparison plots.} Posterior probability distributions of $\delta_{\rm CP}$ (a), $\sin^2 \theta_{23}$ (b), and $\Delta m_{32}^2$ (c) and 1$\sigma$ credible regions in $\Delta m_{32}^2$-$\sin^2 \theta_{23}$ (d), marginalized over both neutrino mass ordering hypotheses (``Both MO'') from fits to pseudo-data simulated with the NuFit-like oscillation parameter values. The fits were run in three configurations while treating the systematic uncertainties with the largest impact on $\sin^2 \theta_{23}$ (visible neutron energy and 2p2h C/O scale) as either 100\% correlated (gray), uncorrelated (teal), or 100\% anticorrelated (magenta). Overlaid with the corresponding 1$\sigma$ (dark shaded areas, dashed) and 2$\sigma$ (light shaded areas, dash-dotted) credible intervals.}
    \label{fig:int_corrs}
\end{figure*}

As an example, Extended Data Fig.~\ref{fig:int_corrs} shows the results in terms of the posterior probability distributions and credible regions of the parameters of interest from the set of fits with the largest single-parameter impact on $\sin^2 \theta_{23}$. We conclude that the choice of correlation between single parameters does not significantly change the oscillation parameter constraints derived from the current version of the joint analysis.

\subsection*{``Nightmare'' Parameters}\label{sec:meth:nightmare}
As described in the main text, we study correlations in more extreme situations using so-called ``nightmare'' parameters, which are either artificially constructed parameters or existing parameters with highly inflated uncertainties chosen to be deliberately problematic for the individual analyses.  The parameters' prior uncertainties are set so that they are comparable in impact to the statistical uncertainties on the measurements under study. We carry out this procedure separately for simulated measurements of $\Delta m^2_{32}$ and $\theta_{23}$. No nightmare study was carried out for $\delta_{\rm CP}$ as its total systematic uncertainty compared to the statistical uncertainty is much smaller than for the other two cases.

\begin{figure}
    \centering
    \includegraphics[width=.5\textwidth]{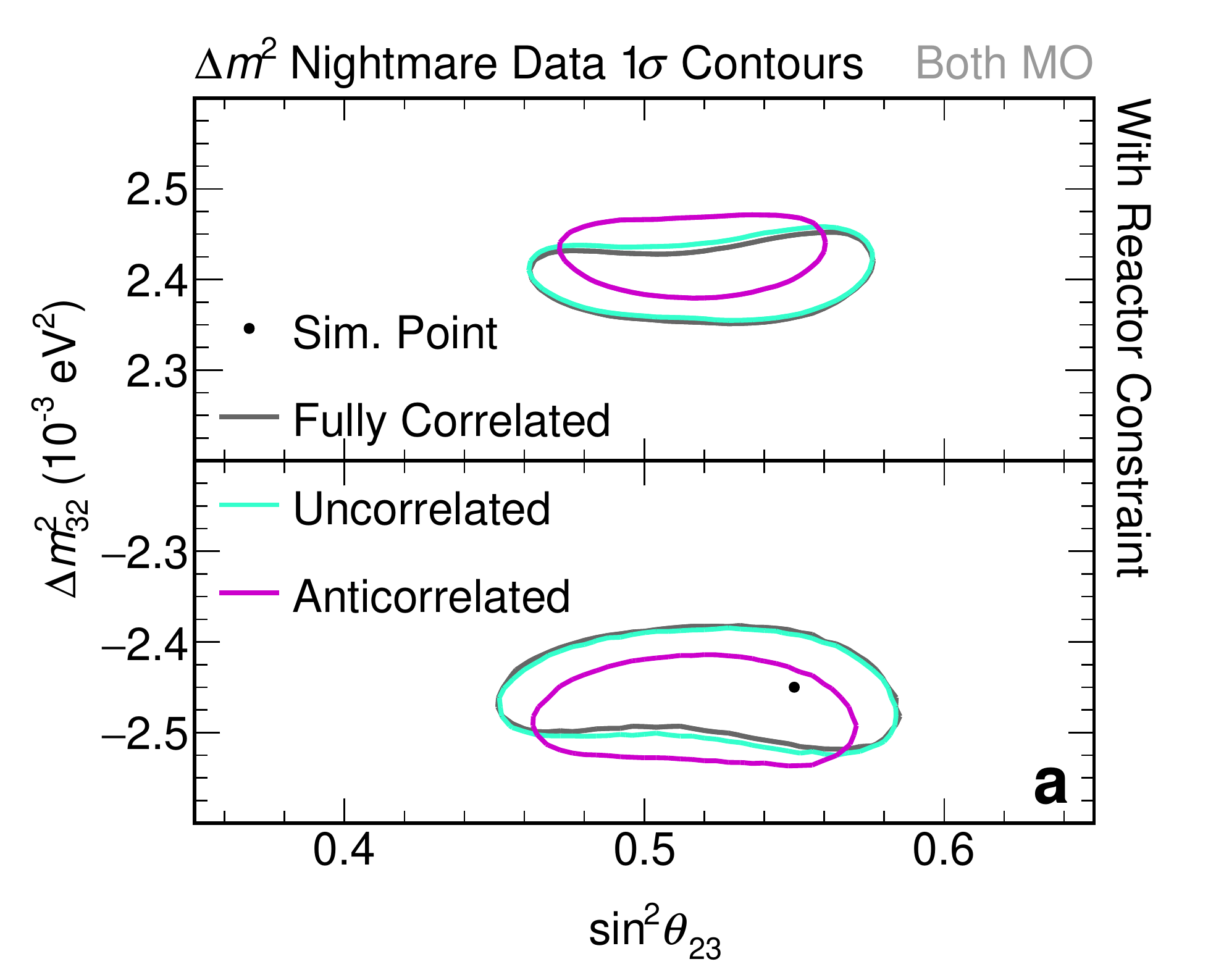}
    \includegraphics[width=.5\textwidth]{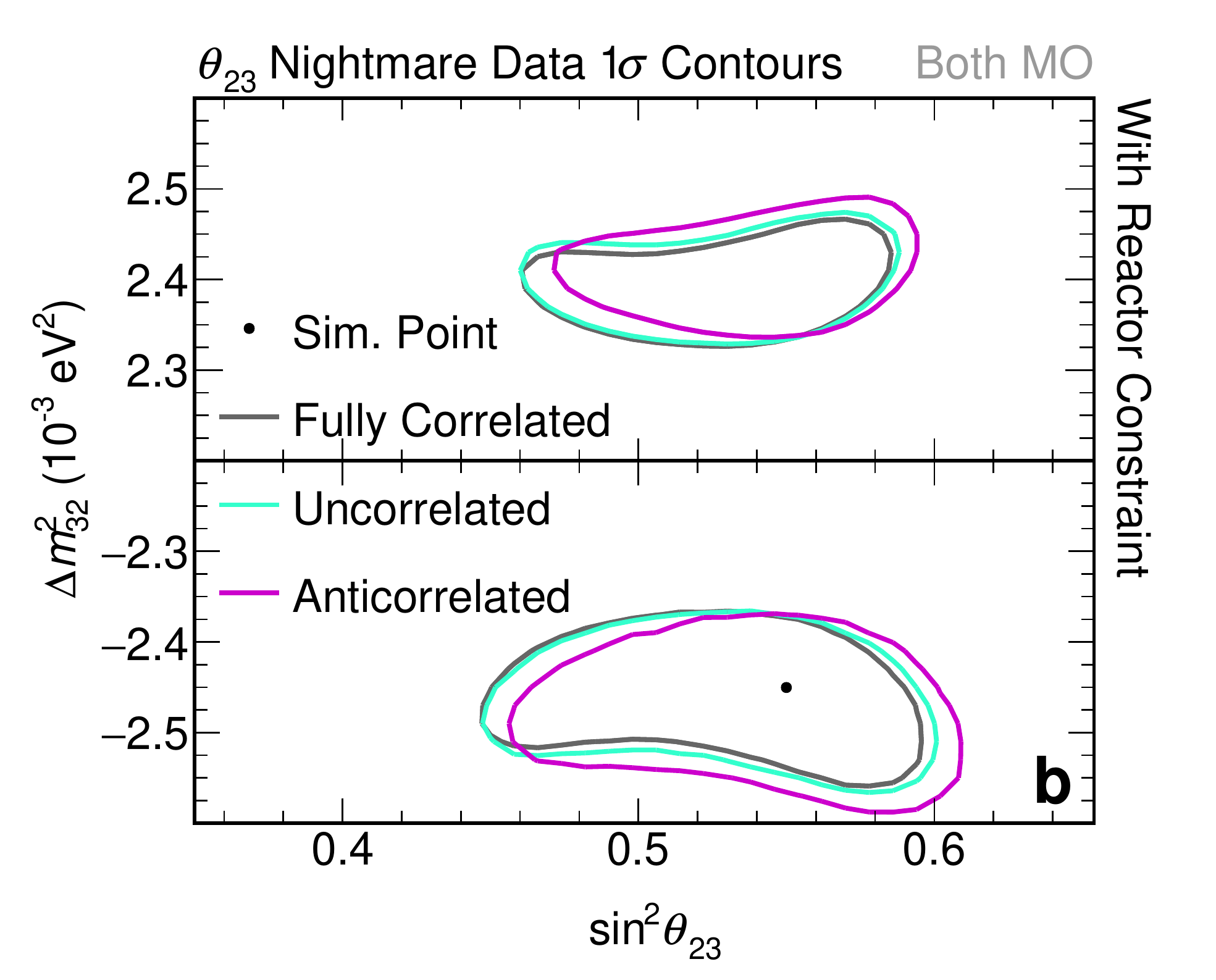}
    \caption{\textbf{``Nightmare'' study comparisons.} 1$\sigma$ credible regions in $\Delta m_{32}^2$-$\sin^2 \theta_{23}$ posterior probability distributions marginalized over both neutrino mass ordering hypotheses (``Both MO'') from fits to pseudo-data simulated with the NuFit-like oscillation parameter values and a fully symmetric systematic bias to affect (a) $\Delta m_{32}^2$ (``$\Delta m^2$ nightmare'') and (b) $\sin^2 \theta_{23}$ (``$\theta_{23}$ nightmare''). The fits were run while treating the NOvA and T2K nightmare parameters as either 100\% correlated (gray), uncorrelated (teal), or 100\% anticorrelated (magenta).}
    \label{fig:nightmare}
\end{figure}

We construct pseudo-data sets with both the NOvA and T2K nightmare parameters shifted by one standard deviation from their prior values, inducing a systematic bias representing a simultaneous and coordinated shift in both NOvA and T2K data. We fit this pseudo-data while treating the NOvA and T2K nightmare parameters as either fully correlated, uncorrelated, or anticorrelated. The results of the nightmare parameters correlation study are presented as $1\sigma$ credible 2D regions of $\Delta m_{32}^2$-$\sin^2 \theta_{23}$ in Extended Data Fig.~\ref{fig:nightmare} for both nightmare scenarios. We conclude that there is no significant difference in treating the nightmare parameters as either fully correlated (matching the pseudo-data) or uncorrelated between the experiments, while the incorrect anticorrelated case yields a clear bias. We note that these are not general conclusions but are specific to the T2K and NOvA analysis versions and cumulative beam exposures used here. The construction of the nightmare parameters is also not a unique choice, and other formulations of the parameters could be considered.

\subsection*{Out-of-Model Variations}\label{sec:meth:fds}
As described in the main text, we use a set of discrete changes to the base cross-section model to test the robustness of our analysis. For each test, pseudo-data are generated assuming the specific model variation, and these pseudo-data are then fit either with the default analysis directly which has no knowledge of the model variation (``out-of-model'' case) or with a modified analysis that has had its nominal event spectra altered to match the spectra expected under the varied model (``in-model’' case). Between these two cases, we require that the width of each of the extracted oscillation parameter intervals changes by no more than 10\% (representing a small ``error on the error'') and that the center of the interval does not move by more than 50\% of the systematic uncertainty (indicating adequate systematic uncertainty coverage of the tested out-of-model variation).  We additionally require that taking the largest changes seen across these studies does not impact the stated conclusions on CP-violation or mass ordering determination for the analysis. 

Three variations were chosen to perform the out-of-model studies.
\begin{itemize}
     \item MINERvA 1$\pi$: This model's suppression of charged current (CC) and neutral current (NC) resonant pion production at low Q$^2$ that was implemented to ensure good agreement between the MINERvA data~\cite{MINERvA:2019kfr} and the implementation of the Rein-Seghal model in the GENIE v2 neutrino interaction simulation software~\cite{Andreopoulos:2009rq}.
      \item Non quasi-elastic (non-QE): In the T2K oscillation analysis~\cite{T2K:2023smv}, the ND280 data samples with a muon candidate and zero pion candidates are underpredicted by the pre-fit T2K nominal model by 10\% in both FGDs, which the fit accounts for by enhancing the charged current quasi-elastic (CCQE) interaction rate. To check this large freedom does not cause bias, an alternate model is produced where this under-prediction is attributed to only non-QE processes.
      \item Pion SI: The pion secondary interaction model in the GEANT4 detector simulation toolkit \cite{GEANT4:2002zbu} was replaced with the Salcedo–Oset model~\cite{Salcedo:1987md} implemented in the NEUT generator~\cite{Hayato:2021heg}, tuned to $\pi$--A scattering data~\cite{PhysRevD.99.052007}.
\end{itemize}

We also used this process to study what happens when fitting pseudo-data constructed for both experiments using one or the other experiment's nominal cross-section model (``T2K-like'' and ``NOvA-like'' studies).

\begin{figure*}
    \centering
    \includegraphics[width=\textwidth]{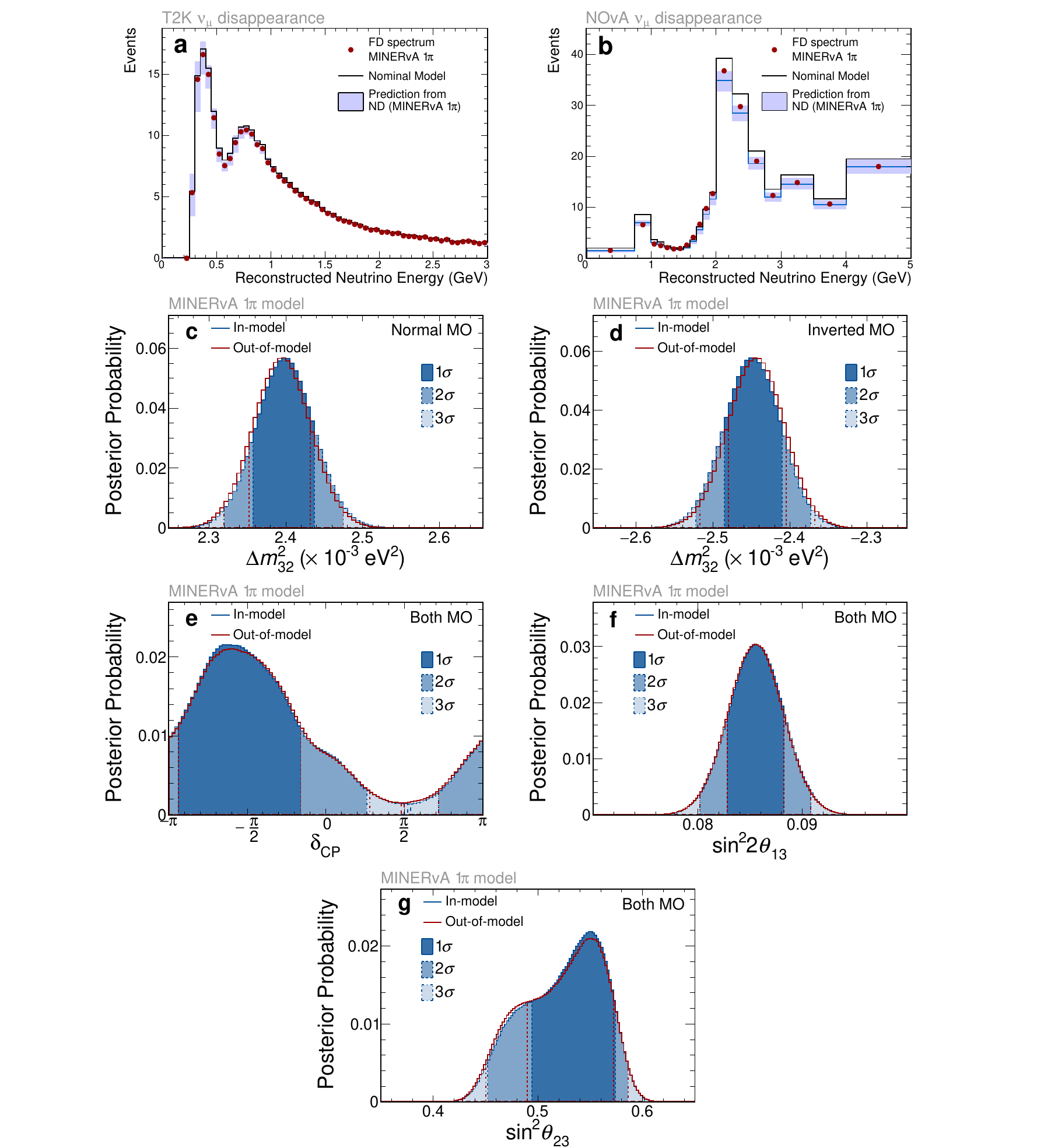}
    \caption{\textbf{Out-of-model study spectra and comparison plots in 1D.} NOvA+T2K out-of-model study with suppressed pion production at low $Q^2$ (``MINERvA 1$\pi$'' case). The change on the FD pseudo-data and prediction with systematic uncertainties after incorporating the alternate data at the ND is shown for for T2K (a) and NOvA (b). Central value of the nominal model is shown for comparison. 1D posterior probability distributions from a fit to pseudo-data generated at the NuFit-like oscillation parameter values are shown for $\Delta m^2_{32}$ marginalized separately over the normal (c) and inverted (d) mass orderings, and for $\delta_{\rm CP}$ (e), $\sin^2 2\theta_{13}$ (f), and $\sin^2 \theta_{23}$ (g) marginalized over both mass orderings. The in-model (blue shaded) and out-of-model (red curve) scenarios are displayed.}

    \label{fig:fds:1d}
\end{figure*}

\begin{figure*}
    \centering
    \includegraphics[width=\textwidth]{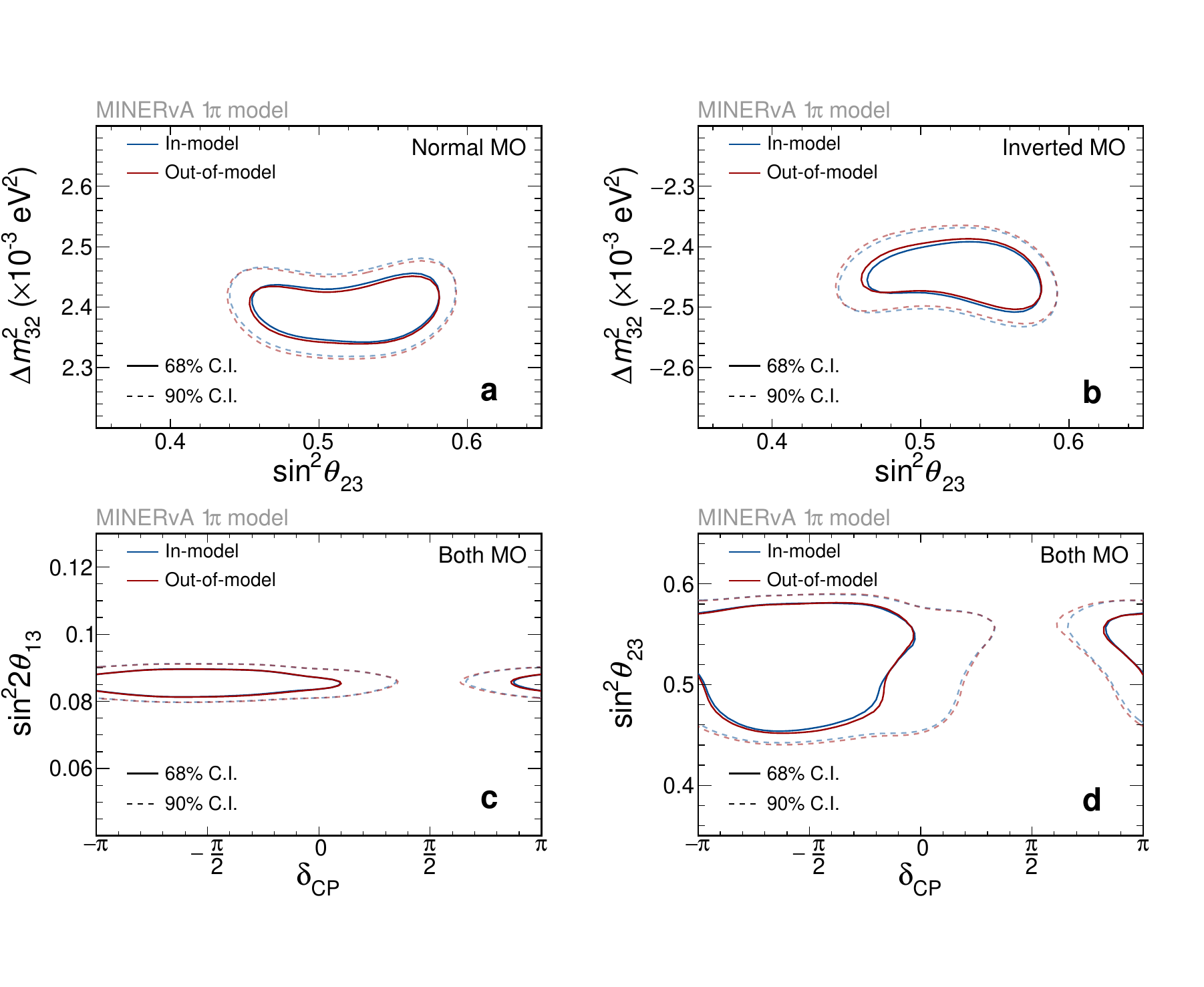}
    \caption{\textbf{Out-of-model study comparison plots in 2D} NOvA+T2K out-of-model study with suppressed pion production at low $Q^2$ (``MINERvA 1$\pi$'' case). 68\% and 90\% contours are shown on the $\sin^2 \theta_{23}$ - $\Delta m^2_{32}$ surface marginalized separately over the normal (a) and inverted (b) mass orderings, and on the surfaces of $\delta_{\rm CP}$-$\sin^2 2\theta_{13}$ (c) and $\delta_{\rm CP}$-$\sin^2 \theta_{23}$ (d) parameters, marginalized over both mass orderings, from a fit to pseudo-data generated at the NuFit-like oscillation parameter values. The in-model (blue shaded) and out-of-model (red curve) scenarios are shown.}
    \label{fig:fds:2d}
\end{figure*}

We show example results here for the MINERvA 1$\pi$ case. Extended Data Fig.~\ref{fig:fds:1d}(a) and (b) illustrate the effect of this alternative model on event spectra used in the analysis (note, not all event spectra are uniformly binned). Extended Data Figs.~\ref{fig:fds:1d}(c)--(g) and \ref{fig:fds:2d} compare the in-model and out-of-model fit results. No failures of our criteria are seen in any of the cases.  More generally, no significant bias is seen in this joint fit for any of the model variations studied across any of the three tested sets of oscillation parameter values.

Some more recent T2K analyses~\cite{abe2024first} did see criteria failures when considering an alternative nuclear model, HF-CRPA~\cite{pandey2015low}, and as a result widened their $\Delta m^2_{32}$ intervals. Both NOvA and T2K have independently studied the impact of the HF-CRPA model on the analyses used in this joint result, and we estimate that any potential effects in the context of this joint fit are within the thresholds set for our out-of-model variation tests.

\subsection*{Goodness-of-Fit}\label{sec:meth:postpred}

The posterior-predictive $p$-value \cite{gellmanpospred} technique is used to determine whether a model provides a good fit to the data it is confronted with. We require that the posterior predictive $p$-value to obtain the far detector data in all samples given the joint post-fit model is greater than 0.05. We also check the $p$-values for individual far detector samples and require that they are greater than 0.05 after allowing for the look-elsewhere effect, using the Bonferroni correction~\cite{Dunn01031961}. All the $p$-values from the joint fit are shown in Extended Data Table~\ref{tab:pppval_both}. All the $p$-values (both total and split sample-by-sample) are within our acceptable range ($>$ 0.05), even without taking the look-elsewhere effect into account. This means that the model used in this joint fit---that is, the individual experiments' systematic models with a shared oscillation parameter model---fits our data well, even when looking at individual samples. The $p$-values are consistent with previous T2K-only and NOvA-only analyses. The $p$-value considering rate and shape for all T2K samples in a T2K-only fit is 0.73, while the $p$-value considering all T2K samples in the joint fit is 0.75. Similarly, the $p$-values for all NOvA samples are 0.56 (NOvA-only fit) and 0.64 (joint fit).

\begin{figure*}
    \centering
    \includegraphics[width=0.45\linewidth]{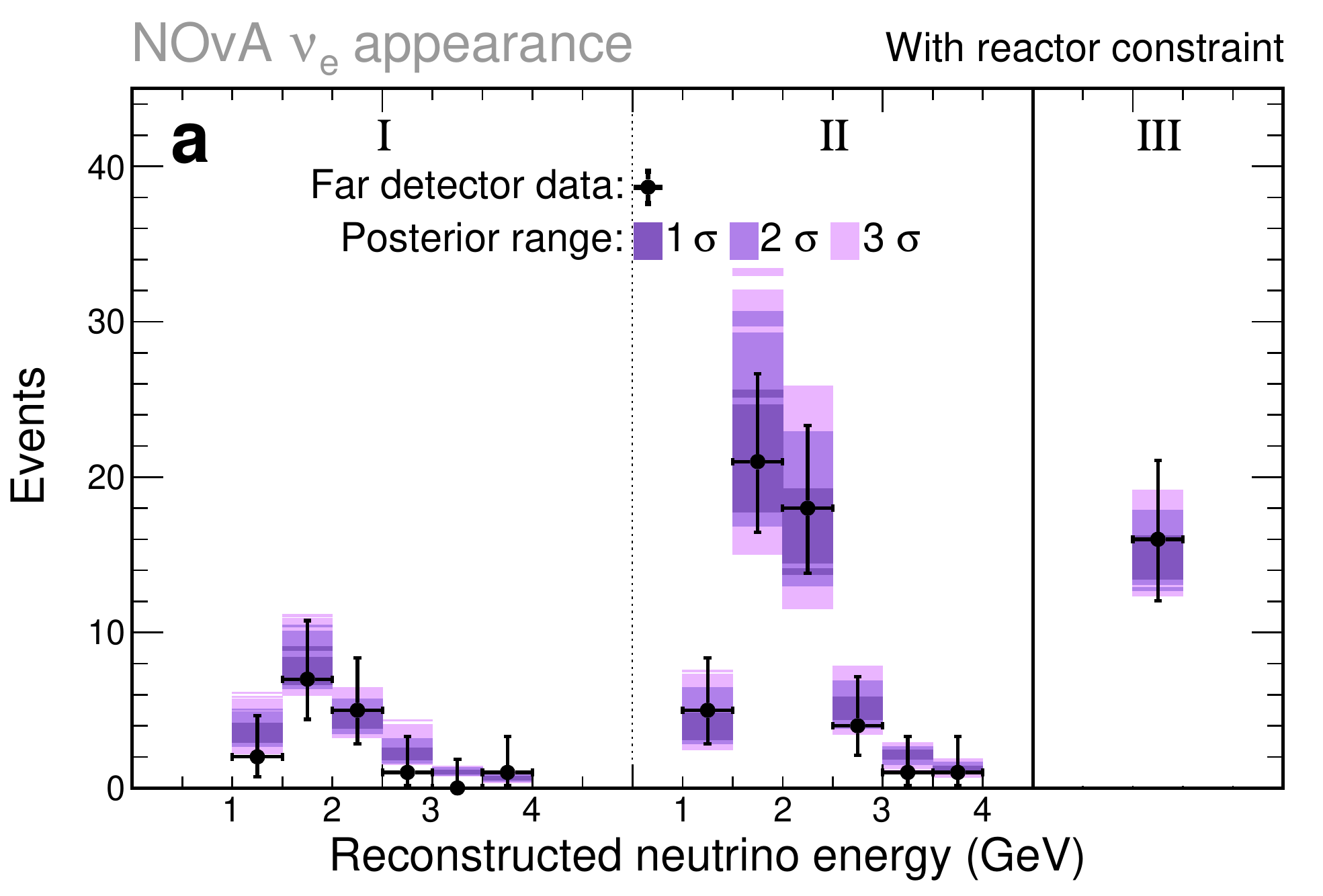}
    \includegraphics[width=0.45\linewidth]{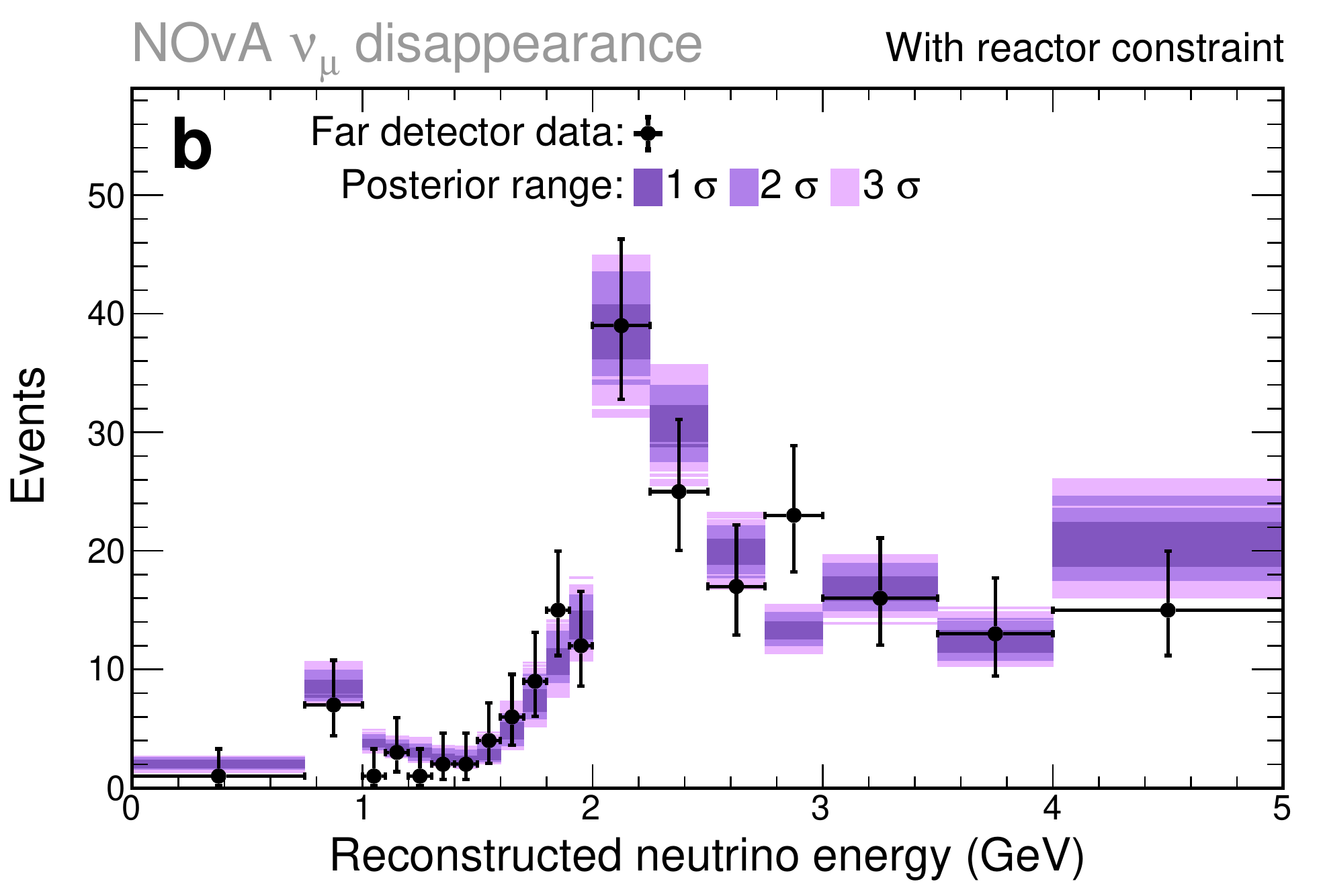}\\
    \includegraphics[width=0.45\linewidth]{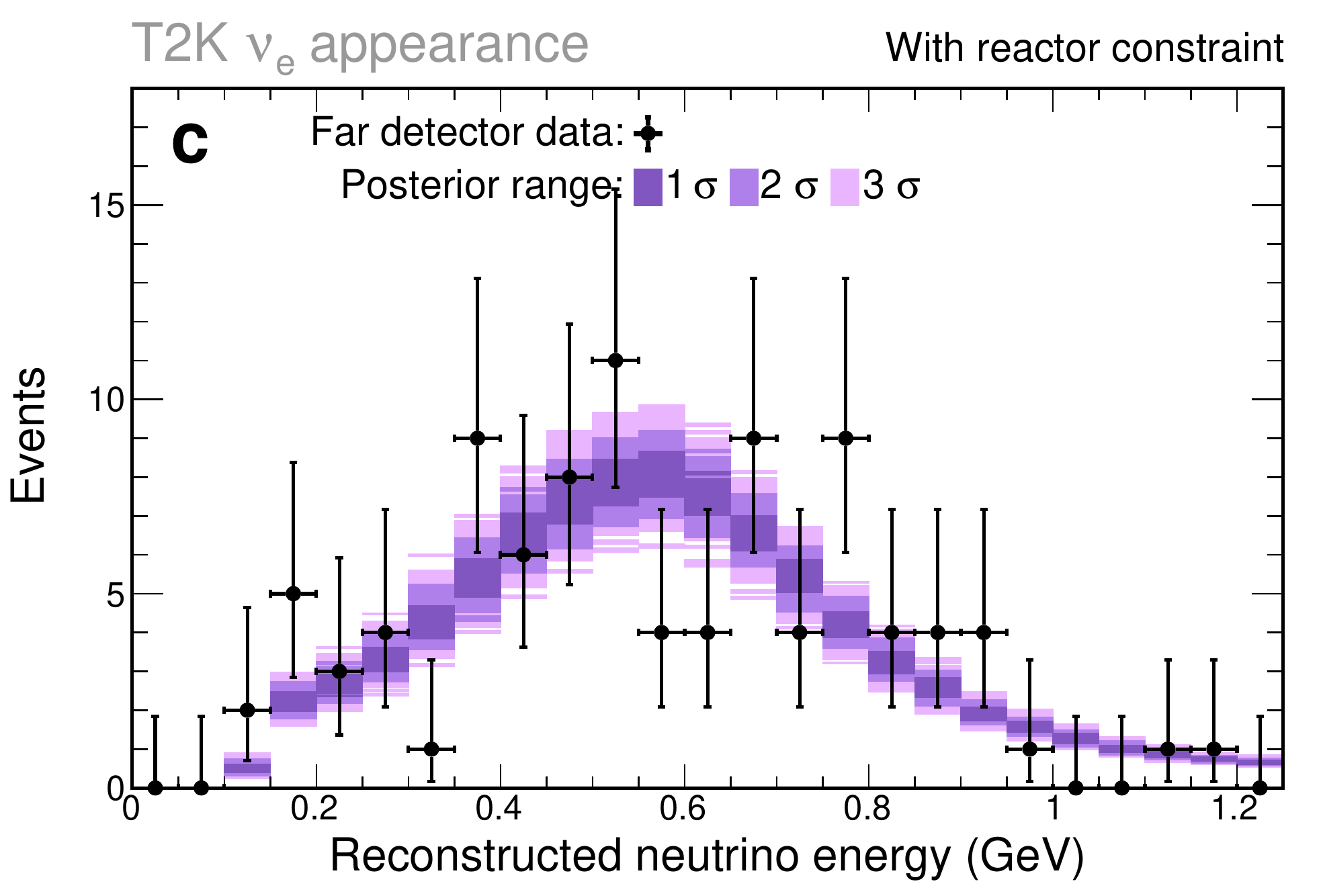}
    \includegraphics[width=0.45\linewidth]{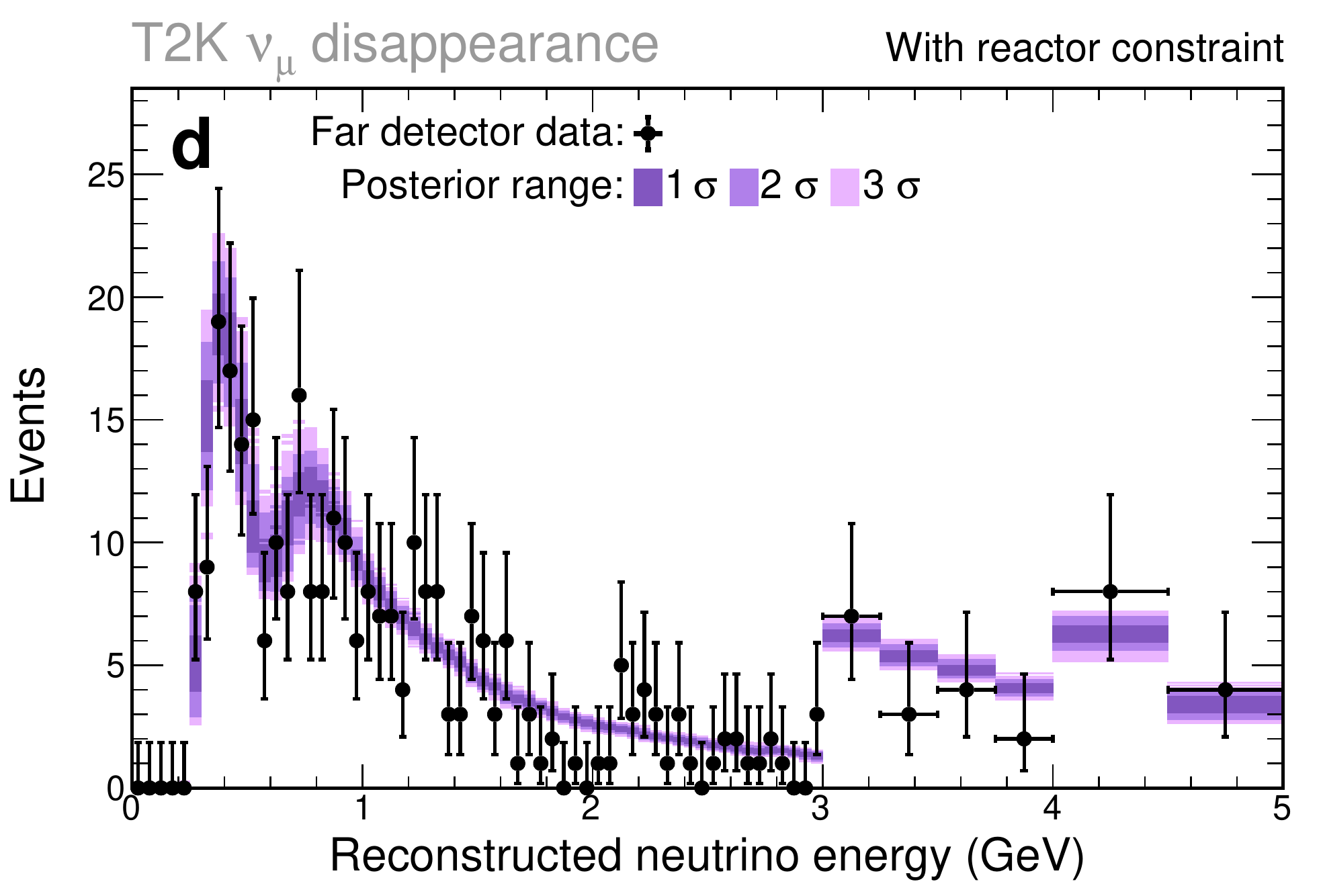}
    \caption{\textbf{NOvA and T2K post-fit spectra.} NOvA~(a, b) and T2K~(c, d) posterior spectra compared to observed data for the largest $\nu_e$-like~(a, c) and $\nu_\mu$-like~(b, d) event samples with the beam running enriched in $\nu_\mu$ (as opposed to $\bar{\nu}_\mu$) extracted from a fit with reactor constraint, marginalized over both mass orderings. The NOvA $\nu_e$-like sample~(a) is divided into three subsets as shown here:\ events with a lower (I) or higher (II) event classification score and events lying near the periphery of the detector (III). Note that T2K also has a $\nu_e$-like sample targeting events with single $\pi$ not shown here.}
    \label{fig:ppval:spec}
\end{figure*}

Example posterior predictions~\cite{gelman_bayesian_2013} of the spectra for both experiments' $\nu_\mu$ and $\nu_e$ subsamples overlaid over the observed data are shown in Extended Data Fig.~\ref{fig:ppval:spec}.

\begin{table}
\caption{\textbf{Posterior predictive \textit{p}-values.} Posterior predictive $p$-values extracted from the joint fits, marginalized over both mass orderings, normal mass ordering and inverted mass ordering with the reactor constraint.}\label{tab:pppval_both}
\vspace{1ex}
    \begin{tabular}{  c  c c c | c c c | c c l }
        \hline
        \multicolumn{10}{c}{Rate + Shape} \\
        \hline
        \multirow{3}{*}{Channel} & \multicolumn{3}{c|}{\multirow{3}{*}{Joint \textit{p}-value}} & \multicolumn{6}{c}{Subsamples \textit{p}-value}\\
                 & & & & \multicolumn{3}{c|}{NOvA\footnote{NOvA: NOvA sample by sample from the joint fit.}} & \multicolumn{3}{c}{T2K\footnote{T2K: T2K sample by sample from the joint fit, $\nu_{e}$ and $\nu_{e}1\pi$ samples treated independently.}} \\
                & Both & NO & IO & Both & NO & IO & Both & NO & IO \\
        \hline
        \multirow{2}{*}{$\sf\nu_e$\footnote{Joint: $\nu_{e}$ channel $p$-value includes T2K $\nu_{e}$, T2K $\nu_{e}1\pi$ and NOvA $\nu_{e}$.}} & \multirow{2}{*}{0.62} & \multirow{2}{*}{0.53} & \multirow{2}{*}{0.69} & \multirow{2}{*}{0.90} & \multirow{2}{*}{0.83} & \multirow{2}{*}{0.95} & 0.19 & 0.18 & $0.20^{(\nu_e)}$ \\
        & & & & & & & 0.79 &  0.78 &  $0.79^{(\nu_e 1\pi)}$\\
        $\bar{\nu}_e$ & 0.40 & 0.38 & 0.42 & 0.21 & 0.18 & 0.24 & 0.67 & 0.67 & 0.67\\
        $\nu_{\mu}$ & 0.62 & 0.62 & 0.62 & 0.68 & 0.65 & 0.70 & 0.48 & 0.50 & 0.47 \\
        $\bar{\nu}_{\mu}$ & 0.72 & 0.73 & 0.71 & 0.38 & 0.38 & 0.37 & 0.87 & 0.87 & 0.87\\
        \hline
        Total & 0.75 & 0.73 & 0.76 & 0.64 & 0.60 & 0.68 & 0.72 & 0.73 & 0.71\\ 
        \hline
        \multicolumn{10}{c}{Rate} \\
        \hline
        \multirow{3}{*}{Channel} & \multicolumn{3}{c|}{\multirow{3}{*}{Joint \textit{p}-value}} & \multicolumn{6}{c}{Subsamples \textit{p}-value}\\
                 & & & & \multicolumn{3}{c|}{NOvA} & \multicolumn{3}{c}{T2K} \\
                & Both & NO & IO & Both & NO & IO & Both & NO & IO \\
        \hline
        $\nu_e$ & 0.40 & 0.14 & 0.57 & 0.48 & 0.16 & 0.71 &  0.39 &  0.43 &  $ 0.36^{(\nu_e)}$ \\ 
        & & & & & & & 0.11 &  0.12 &  $0.11^{(\nu_e 1\pi)}$\\
        $\bar{\nu}_{e}$ & 0.33 & 0.31 & 0.34 & 0.55 & 0.42 & 0.64 & 0.57 & 0.60  & 0.55 \\
        $\nu_{\mu}$ & 0.15 & 0.17 &  0.14 & 0.24 & 0.23 &  0.25 & 0.11 & 0.10 & 0.12\\
        $\bar{\nu}_{\mu}$ & 0.93 & 0.94 & 0.93 & 0.90 & 0.89 & 0.90 & 0.72 & 0.72 & 0.72\\
        \hline
        Total & 0.40 & 0.28  & 0.49 & 0.58 &  0.39 & 0.70 & 0.24 &  0.27 & 0.22 \\ 
        \hline
      \end{tabular}
\end{table}

\subsection*{Priors}
The default priors on the oscillation parameters for this analysis are: flat between $-\pi$ and $\pi$ in $\delta_{\rm{CP}}$, flat between 0 and 1 in $\sin^2\theta_{23}$, flat in $\Delta m^2_{32}$, and Gaussian with $\mu\pm\sigma = (2.18\pm0.007)\times10^{-2}$ in $\sin^2\theta_{13}$. Where alternate priors are used, this is stated in the text. 

This analysis is not sensitive to the oscillation parameters $\sin^2\theta_{12}$ and $\Delta m^2_{21}$ beyond existing experimental constraints; their Gaussian priors are set to be $\sin^2 \theta_{12} = 0.307 \pm 0.013$, and $\Delta m^2_{21} = (7.53 \pm 0.18) \times 10^{-5}~\mbox{eV}^2$. These values, along with a Gaussian prior on $\sin^{2}\theta_{13}$, when it is used, come from the 2020 version of the Particle Data Group (PDG) summary tables~\cite{ParticleDataGroup:2020ssz}, which was current at the time of the original analyses. Updates to these constraints in more recent versions of the PDG do not change any conclusions. 

As well as the standard prior flat in $\delta_{\rm{CP}}$, we also studied the effect of a prior flat in $\sin\delta_{\rm{CP}}$ and saw no significant changes in conclusions.

In addition, the experiments define priors for all of the systematic parameters in their models. These definitions are detailed in the individual experiment analyses underlying this work.

\subsection*{Highest posterior probability values and 1$\sigma$ credible intervals}\label{sec:meth:hpds}
Extended Data Table~\ref{tab:hpds} summarizes the highest posterior probability values and credible intervals measured jointly by NOvA and T2K. 

\begin{table}
\caption{\textbf{NOvA+T2K measurements of oscillation parameters.} Values assume normal and inverted ordering with the reactor constraint applied.}\label{tab:hpds}
\vspace{1ex}
    \centering
    \begin{tabular}{lcc}
    \hline\\[-2ex]
    Parameter & Normal Ordering & Inverted Ordering \\
    \hline\\[-2ex]
    $|\Delta$\textit{m}$^{2}_{32}|$ & $2.43^{+0.04}_{-0.03}$ & $2.48^{+0.04}_{-0.03}$ \\[1ex]
    $\sin^{2}\theta_{23}$ & $0.561^{+0.021}_{-0.039}$ and $0.470^{+0.016}_{-0.008}$\footnote{Local extremum in lower octant of $\sin^{2}\theta_{23}$.} & $0.563^{+0.021}_{-0.039}$\\[1ex]
    $\delta_{ CP}$ & $-0.87\pi^{+0.35\pi}_{-0.21\pi}$ & $-0.47\pi^{+0.17\pi}_{-0.15\pi}$\\[1ex]
    $\sin^{2}2\theta_{13}$ & $ 0.0855\pm 0.0027$ & $0.0859^{+0.0027}_{-0.0025}$\\[.5ex]
    \hline
    \end{tabular}
\end{table}

\subsection*{Additional Oscillation Parameter Plots}\label{sec:meth:oscpar}
\begin{figure*}
    \centering
    \includegraphics[width=\textwidth]{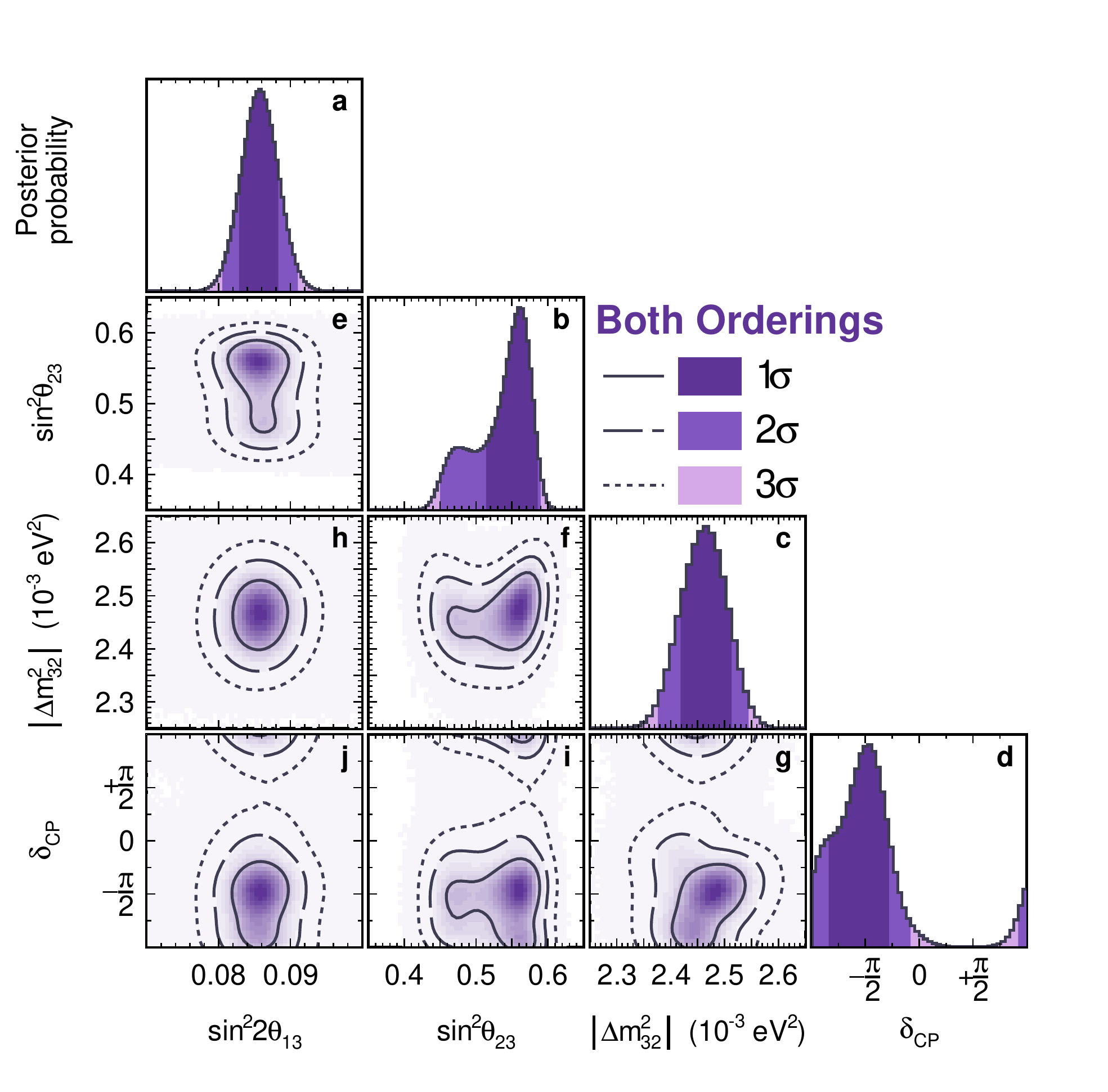}
    \caption{\textbf{Constraints on PMNS oscillation parameters in 1D and 2D for both orderings.} The 1D posterior probability distributions of $\sin^2 2\theta_{13}$ (a), $\sin^2 \theta_{23}$ (b), $|\Delta m^2_{32}|$ (c), $\delta_{\rm CP}$ (d), and corresponding 1$\sigma$, 2$\sigma$, 3$\sigma$ 2D contours $\sin^2\theta_{23}$-$\sin^22\theta_{13}$ (e), $\Delta m_{32}^2$-$\sin^2\theta_{23}$ (f), $\delta_{\rm CP}$-$\Delta m_{32}^2$ (g), $\Delta m_{32}^2$-$\sin^22\theta_{13}$ (h), $\delta_{\rm CP}$-$\sin^2\theta_{23}$ (i), and $\delta_{\rm CP}$-$\sin^22\theta_{13}$ (j) from the joint fit with reactor constraints marginalized over both mass orderings.}
    \label{fig:triangle_both}
\end{figure*}

\begin{figure*}
    \centering
    \includegraphics[width=\textwidth]{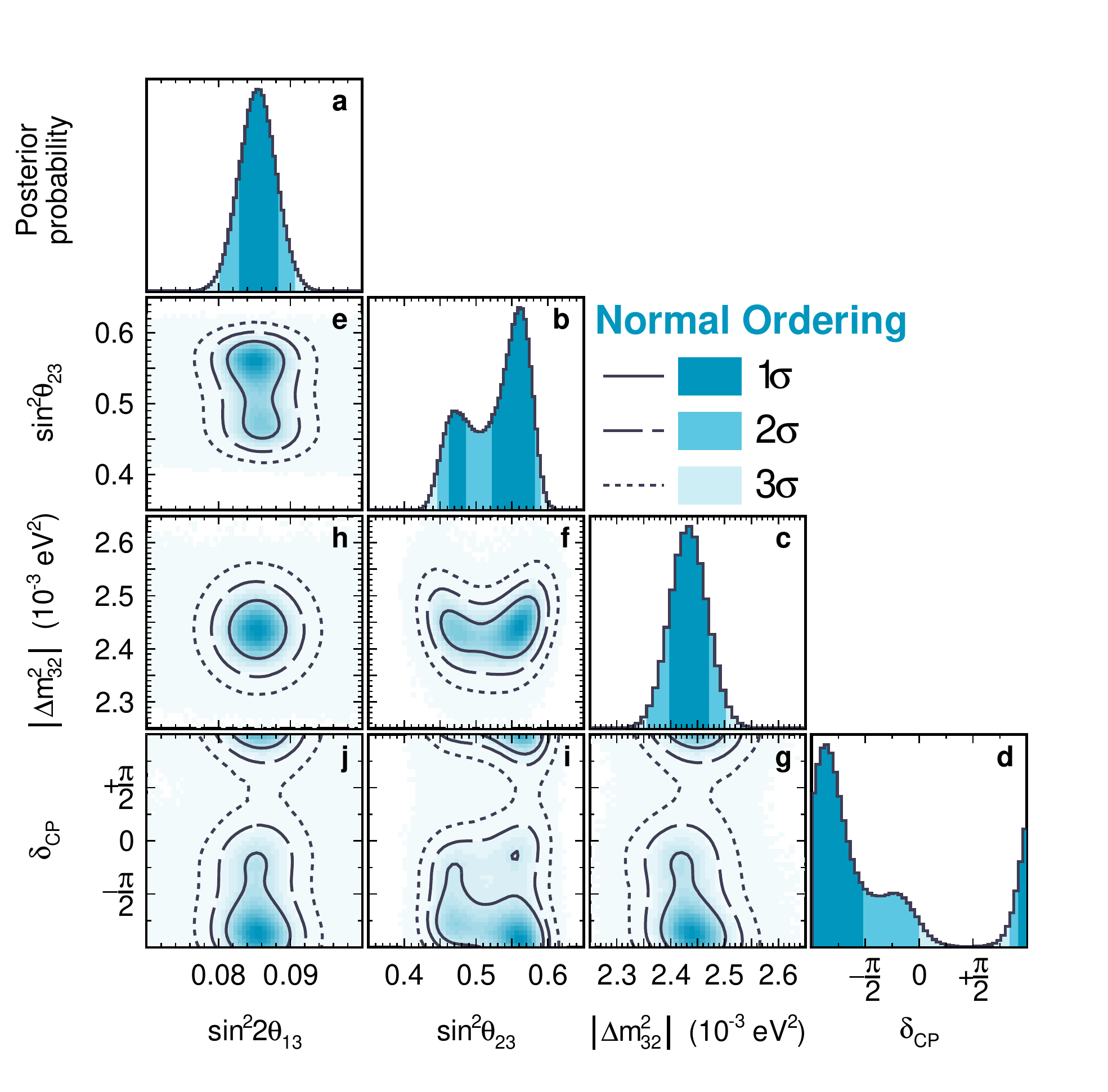}
    \caption{\textbf{Constraints on PMNS oscillation parameters in 1D and 2D for normal ordering.} As in Extended Data Fig.~\ref{fig:triangle_both}, but  conditional on the assumption of normal ordering.}
    \label{fig:triangle_NO}
\end{figure*}

\begin{figure*}
    \centering
    \includegraphics[width=\textwidth]{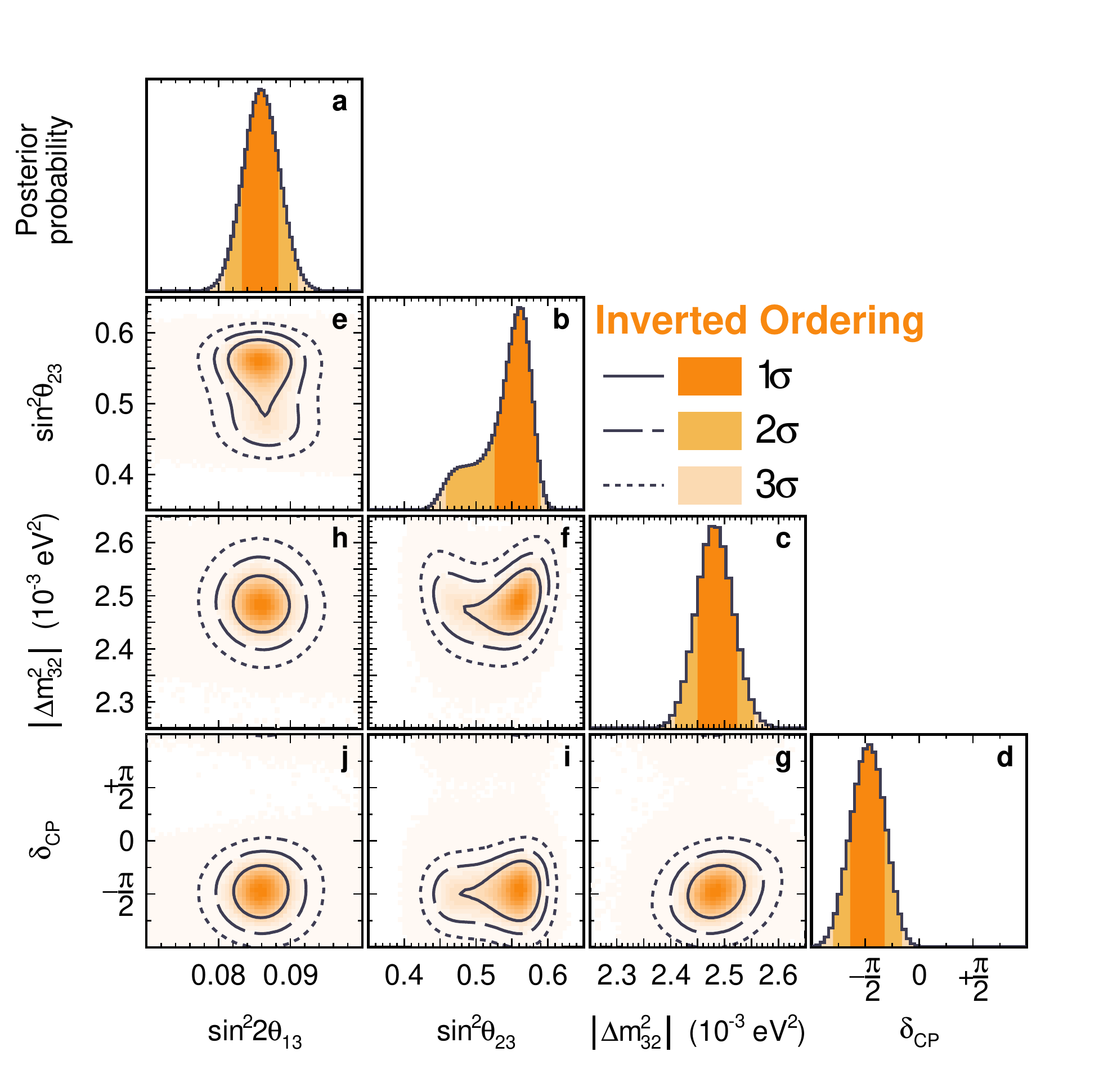}
    \caption{\textbf{Constraints on PMNS oscillation parameters in 1D and 2D for inverted ordering.} As in Extended Data Fig.~\ref{fig:triangle_both}, but conditional on the assumption of inverted ordering.}
    \label{fig:triangle_IO}
\end{figure*}

The main text shows the 1D posterior distributions and credible intervals for the Jarlskog invariant, $\delta_{\rm CP}$, and $\sin^2\theta_{23}$, as well as 2D distributions and credible regions for the latter two. In this section, we present the 1D distributions and credible intervals for $\delta_{\rm CP}$, $\sin^2\theta_{23}$, $\sin^2 2\theta_{13}$, and $|\Delta m^2_{32}|$, and 2D distributions and credible regions for all pairwise combinations of these parameters. These are shown in Extended Data Figs.~\ref{fig:triangle_both},~\ref{fig:triangle_NO}, and~\ref{fig:triangle_IO}, for the cases of marginalized over both mass orderings, conditional on the normal ordering, and conditional on the inverted ordering, respectively. The distributions and intervals are shown in a ``triangle'' plot, where a lower triangular matrix of plots shows the 1D distributions along the diagonal and the 2D distributions in each of the off-diagonal positions. 

\subsection*{Reactor $\Delta m^2_{32}$}\label{sec:meth:2Dreact}
The energy-dependent $\bar{\nu}_e\rightarrow\bar{\nu}_e$ oscillation probability measured by reactor experiments is sensitive to $|\Delta m^2_{32}|$, and reactor measurements of this parameter are expected to agree with long-baseline measurements only under the correct mass ordering assumption.  Under the incorrect ordering assumption, these two techniques are expected to measure incorrect values that differ from one another by $\sim$2--3\%~\cite{Parke:2024xre}. Thus, comparing $|\Delta m^2_{32}|$ measurements from accelerator and reactor experiments under both mass ordering hypotheses can inform mass ordering discrimination. The Daya Bay experiment~\cite{DayaBay:2022orm} provides the tightest constraints on $\theta_{13}$ and also reports a two-dimensional $\theta_{13}$-$\Delta m^2_{32}$ likelihood that we can directly incorporate into our joint fit instead of the $\theta_{13}$-only prior discussed elsewhere in this article.

The mass ordering Bayes factor obtained when using this two-dimensional reactor constraint is 1.4 in favor of the normal ordering, in contrast to 1.3 in favor of the inverted ordering when using the $\theta_{13}$-only reactor constraint.  This slight pull toward a preference for the normal ordering is expected given the relative agreement of the Daya Bay and NOvA+T2K $|\Delta m^2_{32}|$ measurements shown in Fig.~\ref{fig:global_ih} (inverted ordering) and~Extended Data Fig.~\ref{fig:global}(a) (normal ordering.)  However, there remains no statistically significant mass ordering preference in this combination.

\subsection*{Additional Global Comparisons}\label{sec:meth:global}

\begin{figure*}
    \centering
    \includegraphics[width=\textwidth]{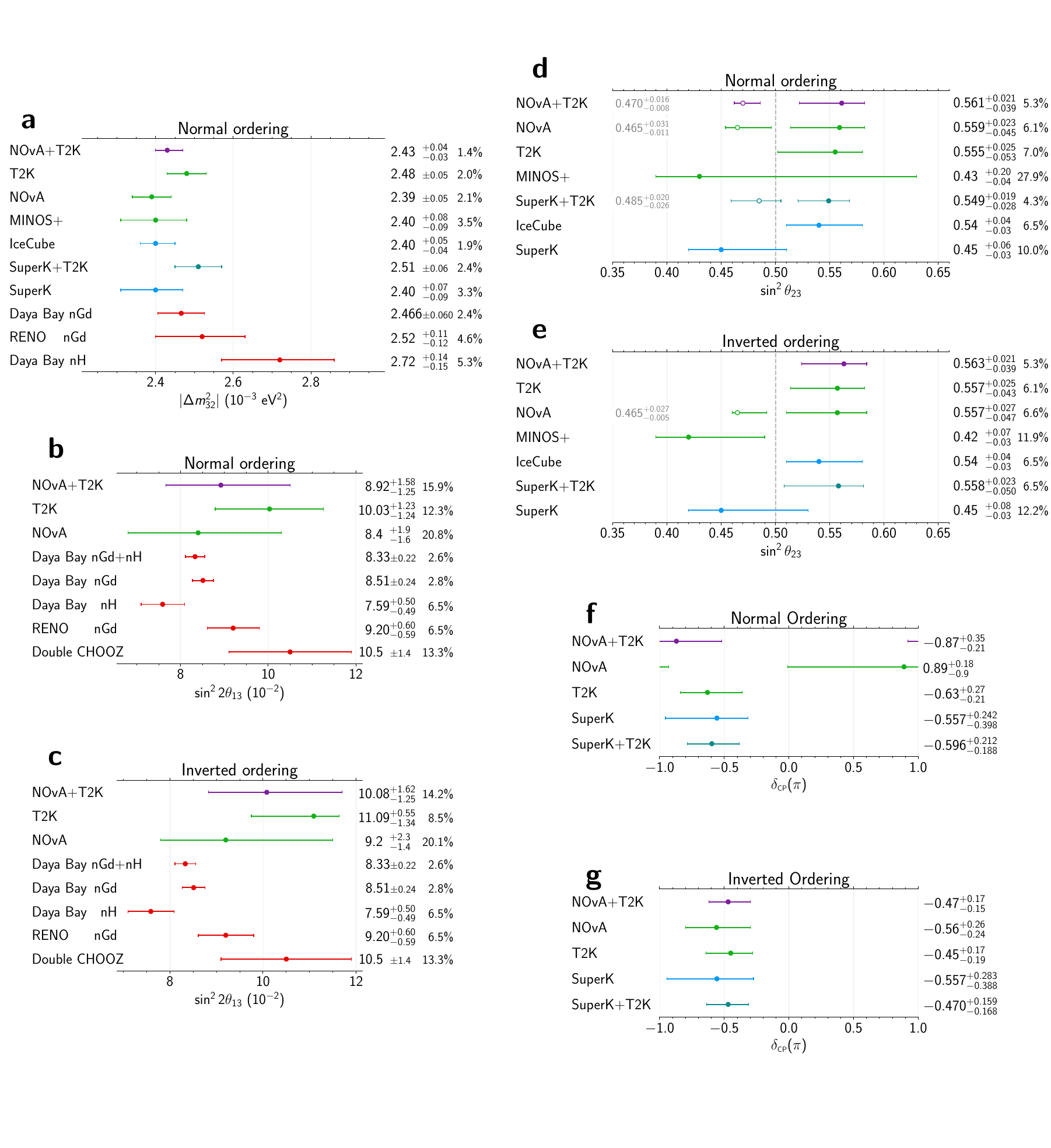}
    \caption{\textbf{Experimental measurements of oscillation parameters} $\vert \Delta m^2_{32}\vert$ assuming normal ordering (a), with sources for the results from top to bottom starting with the second line as follows: \cite{T2K:2023smv, NOvA:2021nfi, MINOS:2020llm, IceCubeCollaboration:2024zec, abe2024first, Super-Kamiokande:2023ahc, DayaBay:2022orm, RENO:2024msr, DayaBay:2024hrv}. $\sin^2 2\theta_{13}$ assuming normal (b) and inverted (c) ordering, with sources for the results from top to bottom starting with the second line as follows: \cite{T2K:2023smv, NOvA:2021nfi,  DayaBay:2024hrv, DayaBay:2022orm,  RENO:2024msr, DoubleChooz:2019qbj}. NOvA+T2K measurement here does not use the reactor constraint. $\sin^2 \theta_{23}$ assuming normal (d) and inverted (e) ordering, with sources for the results from top to bottom starting with the second line as follows: \cite{T2K:2023smv, NOvA:2021nfi, MINOS:2020llm,IceCubeCollaboration:2024zec, abe2024first, Super-Kamiokande:2023ahc}.  Open circles denote a local minima position in lower octant. $\delta_{\rm CP}$ assuming normal (f) and inverted (g) ordering, with sources for the results from top to bottom starting with the second line as follows: \cite{NOvA:2021nfi, T2K:2023smv,  Super-Kamiokande:2023ahc,abe2024first}. }
    \label{fig:global}
\end{figure*}

In Extended Data Fig.~\ref{fig:global} results of the analysis using the default priors are compared with other experimental measurements. The statement on $\Delta m^2_{32}$ precision is still valid for the normal ordering assumption.  As in the case of the $\sin^2 2\theta_{13}$ result (Extended Data Fig.~\ref{fig:global}(b) and (c)), the long-baseline measurements (in this comparison, without applying the prior from reactor measurements) are consistent with reactor experiments, with larger consistency in the normal ordering than the inverted ordering. We do not strongly prefer either octant of $\sin^2 \theta_{23}$ (Extended Data Fig.~\ref{fig:global}(d) and (e)) which is consistent with other modern experiments. The joint analysis $\delta_{\rm CP}$ result (Extended Data Fig.~\ref{fig:global}(f) and (g)) is consistent with all experiments and their combinations although the uncertainty remains large.

\clearpage
\section*{Data Availability}
Inquiries regarding the data and posteriors used in this result may be directed to the collaborations.

\section*{Code Availability}
The NOvA and T2K collaborations develop and maintain the code used for the simulation of the experimental apparatus and statistical analysis of the raw data used in this result.  This code is shared among the collaborations but, due to the size and complexity of the codebases, is not publicly distributed.  Inquiries regarding the algorithms and methods used in this result may be directed to the collaborations.

\section*{Acknowledgements}
The NOvA collaboration thanks Fermi National Accelerator Laboratory (Fermilab), a U.S. Department of Energy, Office of Science, HEP User Facility. Fermilab is managed by Fermi Forward Discovery Group, LLC, acting under Contract No. 89243024CSC000002.  This work was supported by the U.S. Department of Energy, including through the U.S.-Japan Science and Technology Cooperation Program in HEP; the U.S. National Science Foundation; the Department of Science and Technology, India; the European Research Council; the MSMT CR, GA UK, Czech Republic; the RAS, the Ministry of Science and Higher Education, and RFBR, Russia; CNPq and FAPEG, Brazil; UKRI, STFC and the Royal Society, United Kingdom; and the state and University of Minnesota.  We are grateful for the contributions of the staffs of the University of Minnesota at the Ash River Laboratory, and of Fermilab.

The T2K collaboration would like to thank the J-PARC staff for superb accelerator performance. We thank the CERN NA61/SHINE Collaboration for providing valuable particle production data. We acknowledge the support of MEXT,   JSPS KAKENHI  and bilateral programs, Japan; NSERC, the NRC, and CFI, Canada; the CEA and CNRS/IN2P3, France; the Deutsche Forschungsgemeinschaft (DFG 397763730, 517206441), Germany; the NKFIH  (NKFIH 137812 and TKP2021-NKTA-64), Hungary; the INFN, Italy; the Ministry of Science and Higher Education (2023/WK/04) and the National Science Centre (UMO-2018/30/E/ST2/00441, UMO-2022/46/E/ST2/00336 and UMO-2021/43/D/ST2/01504), Poland; the RSF (RSF 24-12-00271) and the Ministry of Science and Higher Education, Russia; MICINN, ERDF and European Union NextGenerationEU funds and CERCA program and Generalitat de Catalunya, Spain; the SNSF and SERI, Switzerland; the STFC and UKRI, UK; the DOE, USA, including through the U.S.-Japan Science and Technology Cooperation Program in HEP; and NAFOSTED (103.99-2023.144,IZVSZ2.203433), Vietnam. We also thank CERN for the UA1/NOMAD magnet, DESY for the HERA-B magnet mover system, the BC DRI Group, Prairie DRI Group, ACENET, SciNet, and CalculQuebec consortia in the Digital Research Alliance of Canada, and GridPP in the United Kingdom, the CNRS/IN2P3 Computing Center in France and NERSC, USA. In addition, the participation of individual researchers and institutions has been further supported by funds from the ERC (FP7), “la Caixa” Foundation, the European Union’s Horizon 2020 Research and Innovation Programme under the Marie Sklodowska-Curie grant; the JSPS, Japan; the Royal Society, UK; French ANR and Sorbonne Université Emergences programmes; the VAST-JSPS (No.\ QTJP01.02/20-22);  and the DOE Early Career program, USA.

For the purpose of open access, the authors have applied a Creative Commons Attribution (CC BY) license to any Author Accepted Manuscript version arising.

\section*{Author contributions}
The operation, Monte Carlo simulation, and data analysis of the T2K and NOvA experiments are carried out by the T2K and NOvA Collaborations with contributions from all collaborators listed as authors on this manuscript.  The scientific results presented here have been presented to and discussed by the full collaborations, and all authors have approved the final version of the manuscript.

\section*{Competing interests}
The authors declare no competing interests.

\section*{Additional Information}
Correspondence and requests for materials should be addressed to the T2K and NOvA collaborations. 

\bibliography{cites}

\begin{thebibliography}{63}%
\makeatletter
\providecommand \@ifxundefined [1]{%
 \@ifx{#1\undefined}
}%
\providecommand \@ifnum [1]{%
 \ifnum #1\expandafter \@firstoftwo
 \else \expandafter \@secondoftwo
 \fi
}%
\providecommand \@ifx [1]{%
 \ifx #1\expandafter \@firstoftwo
 \else \expandafter \@secondoftwo
 \fi
}%
\providecommand \natexlab [1]{#1}%
\providecommand \enquote  [1]{``#1''}%
\providecommand \bibnamefont  [1]{#1}%
\providecommand \bibfnamefont [1]{#1}%
\providecommand \citenamefont [1]{#1}%
\providecommand \href@noop [0]{\@secondoftwo}%
\providecommand \href [0]{\begingroup \@sanitize@url \@href}%
\providecommand \@href[1]{\@@startlink{#1}\@@href}%
\providecommand \@@href[1]{\endgroup#1\@@endlink}%
\providecommand \@sanitize@url [0]{\catcode `\\12\catcode `\$12\catcode `\&12\catcode `\#12\catcode `\^12\catcode `\_12\catcode `\%12\relax}%
\providecommand \@@startlink[1]{}%
\providecommand \@@endlink[0]{}%
\providecommand \url  [0]{\begingroup\@sanitize@url \@url }%
\providecommand \@url [1]{\endgroup\@href {#1}{\urlprefix }}%
\providecommand \urlprefix  [0]{URL }%
\providecommand \Eprint [0]{\href }%
\providecommand \doibase [0]{https://doi.org/}%
\providecommand \selectlanguage [0]{\@gobble}%
\providecommand \bibinfo  [0]{\@secondoftwo}%
\providecommand \bibfield  [0]{\@secondoftwo}%
\providecommand \translation [1]{[#1]}%
\providecommand \BibitemOpen [0]{}%
\providecommand \bibitemStop [0]{}%
\providecommand \bibitemNoStop [0]{.\EOS\space}%
\providecommand \EOS [0]{\spacefactor3000\relax}%
\providecommand \BibitemShut  [1]{\csname bibitem#1\endcsname}%
\let\auto@bib@innerbib\@empty
\bibitem [{\citenamefont {Fukuda}\ \emph {et~al.}(1998)\citenamefont {Fukuda} \emph {et~al.}}]{Super-Kamiokande:1998kpq}%
  \BibitemOpen
  \bibfield  {author} {\bibinfo {author} {\bibfnamefont {Y.}~\bibnamefont {Fukuda}} \emph {et~al.} (\bibinfo {collaboration} {Super-Kamiokande}),\ }\bibfield  {title} {\bibinfo {title} {{Evidence for oscillation of atmospheric neutrinos}},\ }\href {https://doi.org/10.1103/PhysRevLett.81.1562} {\bibfield  {journal} {\bibinfo  {journal} {Phys. Rev. Lett.}\ }\textbf {\bibinfo {volume} {81}},\ \bibinfo {pages} {1562} (\bibinfo {year} {1998})}\BibitemShut {NoStop}%
\bibitem [{\citenamefont {Fukuda}\ \emph {et~al.}(2002)\citenamefont {Fukuda} \emph {et~al.}}]{Super-Kamiokande:2002ujc}%
  \BibitemOpen
  \bibfield  {author} {\bibinfo {author} {\bibfnamefont {S.}~\bibnamefont {Fukuda}} \emph {et~al.} (\bibinfo {collaboration} {Super-Kamiokande}),\ }\bibfield  {title} {\bibinfo {title} {{Determination of solar neutrino oscillation parameters using 1496 days of Super-Kamiokande I data}},\ }\href {https://doi.org/10.1016/S0370-2693(02)02090-7} {\bibfield  {journal} {\bibinfo  {journal} {Phys. Lett. B}\ }\textbf {\bibinfo {volume} {539}},\ \bibinfo {pages} {179} (\bibinfo {year} {2002})}\BibitemShut {NoStop}%
\bibitem [{\citenamefont {Ahmad}\ \emph {et~al.}(2001)\citenamefont {Ahmad} \emph {et~al.}}]{SNO:2001kpb}%
  \BibitemOpen
  \bibfield  {author} {\bibinfo {author} {\bibfnamefont {Q.~R.}\ \bibnamefont {Ahmad}} \emph {et~al.} (\bibinfo {collaboration} {SNO}),\ }\bibfield  {title} {\bibinfo {title} {{Measurement of the rate of $\nu_e+d \to p+p+e^-$ interactions produced by $^8$B solar neutrinos at the Sudbury Neutrino Observatory}},\ }\href {https://doi.org/10.1103/PhysRevLett.87.071301} {\bibfield  {journal} {\bibinfo  {journal} {Phys. Rev. Lett.}\ }\textbf {\bibinfo {volume} {87}},\ \bibinfo {pages} {071301} (\bibinfo {year} {2001})}\BibitemShut {NoStop}%
\bibitem [{\citenamefont {Ahmad}\ \emph {et~al.}(2002)\citenamefont {Ahmad} \emph {et~al.}}]{SNO:2002tuh}%
  \BibitemOpen
  \bibfield  {author} {\bibinfo {author} {\bibfnamefont {Q.~R.}\ \bibnamefont {Ahmad}} \emph {et~al.} (\bibinfo {collaboration} {SNO}),\ }\bibfield  {title} {\bibinfo {title} {{Direct evidence for neutrino flavor transformation from neutral current interactions in the Sudbury Neutrino Observatory}},\ }\href {https://doi.org/10.1103/PhysRevLett.89.011301} {\bibfield  {journal} {\bibinfo  {journal} {Phys. Rev. Lett.}\ }\textbf {\bibinfo {volume} {89}},\ \bibinfo {pages} {011301} (\bibinfo {year} {2002})}\BibitemShut {NoStop}%
\bibitem [{\citenamefont {Eguchi}\ \emph {et~al.}(2003)\citenamefont {Eguchi} \emph {et~al.}}]{KamLAND:2002uet}%
  \BibitemOpen
  \bibfield  {author} {\bibinfo {author} {\bibfnamefont {K.}~\bibnamefont {Eguchi}} \emph {et~al.} (\bibinfo {collaboration} {KamLAND}),\ }\bibfield  {title} {\bibinfo {title} {{First results from KamLAND: Evidence for reactor anti-neutrino disappearance}},\ }\href {https://doi.org/10.1103/PhysRevLett.90.021802} {\bibfield  {journal} {\bibinfo  {journal} {Phys. Rev. Lett.}\ }\textbf {\bibinfo {volume} {90}},\ \bibinfo {pages} {021802} (\bibinfo {year} {2003})}\BibitemShut {NoStop}%
\bibitem [{\citenamefont {An}\ \emph {et~al.}(2012)\citenamefont {An} \emph {et~al.}}]{DayaBay:2012fng}%
  \BibitemOpen
  \bibfield  {author} {\bibinfo {author} {\bibfnamefont {F.~P.}\ \bibnamefont {An}} \emph {et~al.} (\bibinfo {collaboration} {Daya Bay}),\ }\bibfield  {title} {\bibinfo {title} {{Observation of electron-antineutrino disappearance at Daya Bay}},\ }\href {https://doi.org/10.1103/PhysRevLett.108.171803} {\bibfield  {journal} {\bibinfo  {journal} {Phys. Rev. Lett.}\ }\textbf {\bibinfo {volume} {108}},\ \bibinfo {pages} {171803} (\bibinfo {year} {2012})}\BibitemShut {NoStop}%
\bibitem [{\citenamefont {Fukugita}\ and\ \citenamefont {Yanagida}(1986)}]{Fukugita:1986hr}%
  \BibitemOpen
  \bibfield  {author} {\bibinfo {author} {\bibfnamefont {M.}~\bibnamefont {Fukugita}}\ and\ \bibinfo {author} {\bibfnamefont {T.}~\bibnamefont {Yanagida}},\ }\bibfield  {title} {\bibinfo {title} {{Baryogenesis without grand unification}},\ }\href {https://doi.org/10.1016/0370-2693(86)91126-3} {\bibfield  {journal} {\bibinfo  {journal} {Phys. Lett. B}\ }\textbf {\bibinfo {volume} {174}},\ \bibinfo {pages} {45} (\bibinfo {year} {1986})}\BibitemShut {NoStop}%
\bibitem [{\citenamefont {Buchmuller}\ \emph {et~al.}(2005)\citenamefont {Buchmuller}, \citenamefont {Peccei},\ and\ \citenamefont {Yanagida}}]{Buchmuller:2005eh}%
  \BibitemOpen
  \bibfield  {author} {\bibinfo {author} {\bibfnamefont {W.}~\bibnamefont {Buchmuller}}, \bibinfo {author} {\bibfnamefont {R.~D.}\ \bibnamefont {Peccei}},\ and\ \bibinfo {author} {\bibfnamefont {T.}~\bibnamefont {Yanagida}},\ }\bibfield  {title} {\bibinfo {title} {{Leptogenesis as the origin of matter}},\ }\href {https://doi.org/10.1146/annurev.nucl.55.090704.151558} {\bibfield  {journal} {\bibinfo  {journal} {Ann. Rev. Nucl. Part. Sci.}\ }\textbf {\bibinfo {volume} {55}},\ \bibinfo {pages} {311} (\bibinfo {year} {2005})},\ \Eprint {https://arxiv.org/abs/hep-ph/0502169} {arXiv:hep-ph/0502169} \BibitemShut {NoStop}%
\bibitem [{\citenamefont {Pascoli}\ \emph {et~al.}(2007)\citenamefont {Pascoli}, \citenamefont {Petcov},\ and\ \citenamefont {Riotto}}]{pascoli2007connecting}%
  \BibitemOpen
  \bibfield  {author} {\bibinfo {author} {\bibfnamefont {S.}~\bibnamefont {Pascoli}}, \bibinfo {author} {\bibfnamefont {S.~T.}\ \bibnamefont {Petcov}},\ and\ \bibinfo {author} {\bibfnamefont {A.}~\bibnamefont {Riotto}},\ }\bibfield  {title} {\bibinfo {title} {{Connecting low energy leptonic CP violation to leptogenesis}},\ }\href {https://doi.org/10.1103/PhysRevD.75.083511} {\bibfield  {journal} {\bibinfo  {journal} {Phys. Rev. D}\ }\textbf {\bibinfo {volume} {75}},\ \bibinfo {pages} {083511} (\bibinfo {year} {2007})}\BibitemShut {NoStop}%
\bibitem [{\citenamefont {Branco}\ \emph {et~al.}(2012)\citenamefont {Branco}, \citenamefont {Gonzalez~Felipe},\ and\ \citenamefont {Joaquim}}]{branco2012leptonic}%
  \BibitemOpen
  \bibfield  {author} {\bibinfo {author} {\bibfnamefont {G.~C.}\ \bibnamefont {Branco}}, \bibinfo {author} {\bibfnamefont {R.}~\bibnamefont {Gonzalez~Felipe}},\ and\ \bibinfo {author} {\bibfnamefont {F.~R.}\ \bibnamefont {Joaquim}},\ }\bibfield  {title} {\bibinfo {title} {{Leptonic CP violation}},\ }\href {https://doi.org/10.1103/RevModPhys.84.515} {\bibfield  {journal} {\bibinfo  {journal} {Rev. Mod. Phys.}\ }\textbf {\bibinfo {volume} {84}},\ \bibinfo {pages} {515} (\bibinfo {year} {2012})}\BibitemShut {NoStop}%
\bibitem [{\citenamefont {Hagedorn}\ \emph {et~al.}(2018)\citenamefont {Hagedorn}, \citenamefont {Mohapatra}, \citenamefont {Molinaro}, \citenamefont {Nishi},\ and\ \citenamefont {Petcov}}]{hagedorn2018cp}%
  \BibitemOpen
  \bibfield  {author} {\bibinfo {author} {\bibfnamefont {C.}~\bibnamefont {Hagedorn}}, \bibinfo {author} {\bibfnamefont {R.}~\bibnamefont {Mohapatra}}, \bibinfo {author} {\bibfnamefont {E.}~\bibnamefont {Molinaro}}, \bibinfo {author} {\bibfnamefont {C.}~\bibnamefont {Nishi}},\ and\ \bibinfo {author} {\bibfnamefont {S.}~\bibnamefont {Petcov}},\ }\bibfield  {title} {\bibinfo {title} {{CP violation in the lepton sector and implications for leptogenesis}},\ }\href {https://doi.org/10.1142/S0217751X1842006X} {\bibfield  {journal} {\bibinfo  {journal} {Int. J. Mod. Phys. A}\ }\textbf {\bibinfo {volume} {33}},\ \bibinfo {pages} {1842006} (\bibinfo {year} {2018})}\BibitemShut {NoStop}%
\bibitem [{\citenamefont {Workman}\ \emph {et~al.}(2022)\citenamefont {Workman} \emph {et~al.}}]{ParticleDataGroup:2022pth}%
  \BibitemOpen
  \bibfield  {author} {\bibinfo {author} {\bibfnamefont {R.~L.}\ \bibnamefont {Workman}} \emph {et~al.} (\bibinfo {collaboration} {Particle Data Group}),\ }\bibfield  {title} {\bibinfo {title} {{Review of particle physics}},\ }\href {https://doi.org/10.1093/ptep/ptac097} {\bibfield  {journal} {\bibinfo  {journal} {PTEP}\ }\textbf {\bibinfo {volume} {2022}},\ \bibinfo {pages} {083C01} (\bibinfo {year} {2022})}\BibitemShut {NoStop}%
\bibitem [{\citenamefont {Acero}\ \emph {et~al.}(2022)\citenamefont {Acero} \emph {et~al.}}]{NOvA:2021nfi}%
  \BibitemOpen
  \bibfield  {author} {\bibinfo {author} {\bibfnamefont {M.~A.}\ \bibnamefont {Acero}} \emph {et~al.} (\bibinfo {collaboration} {NOvA}),\ }\bibfield  {title} {\bibinfo {title} {{Improved measurement of neutrino oscillation parameters by the NOvA experiment}},\ }\href {https://doi.org/10.1103/PhysRevD.106.032004} {\bibfield  {journal} {\bibinfo  {journal} {Phys. Rev. D}\ }\textbf {\bibinfo {volume} {106}},\ \bibinfo {pages} {032004} (\bibinfo {year} {2022})}\BibitemShut {NoStop}%
\bibitem [{\citenamefont {Abe}\ \emph {et~al.}(2023)\citenamefont {Abe} \emph {et~al.}}]{T2K:2023smv}%
  \BibitemOpen
  \bibfield  {author} {\bibinfo {author} {\bibfnamefont {K.}~\bibnamefont {Abe}} \emph {et~al.} (\bibinfo {collaboration} {T2K}),\ }\bibfield  {title} {\bibinfo {title} {{Measurements of neutrino oscillation parameters from the T2K experiment using $3.6\times 10^{21}$ protons on target}},\ }\href {https://doi.org/10.1140/epjc/s10052-023-11819-x} {\bibfield  {journal} {\bibinfo  {journal} {Eur. Phys. J. C}\ }\textbf {\bibinfo {volume} {83}},\ \bibinfo {pages} {782} (\bibinfo {year} {2023})}\BibitemShut {NoStop}%
\bibitem [{\citenamefont {Maki}\ \emph {et~al.}(1962)\citenamefont {Maki}, \citenamefont {Nakagawa},\ and\ \citenamefont {Sakata}}]{Maki:1962mu}%
  \BibitemOpen
  \bibfield  {author} {\bibinfo {author} {\bibfnamefont {Z.}~\bibnamefont {Maki}}, \bibinfo {author} {\bibfnamefont {M.}~\bibnamefont {Nakagawa}},\ and\ \bibinfo {author} {\bibfnamefont {S.}~\bibnamefont {Sakata}},\ }\bibfield  {title} {\bibinfo {title} {{Remarks on the unified model of elementary particles}},\ }\href {https://doi.org/10.1143/PTP.28.870} {\bibfield  {journal} {\bibinfo  {journal} {Prog. Theor. Phys.}\ }\textbf {\bibinfo {volume} {28}},\ \bibinfo {pages} {870} (\bibinfo {year} {1962})}\BibitemShut {NoStop}%
\bibitem [{\citenamefont {Pontecorvo}(1967)}]{Pontecorvo:1967fh}%
  \BibitemOpen
  \bibfield  {author} {\bibinfo {author} {\bibfnamefont {B.}~\bibnamefont {Pontecorvo}},\ }\bibfield  {title} {\bibinfo {title} {{Neutrino experiments and the problem of conservation of leptonic charge}},\ }\href@noop {} {\bibfield  {journal} {\bibinfo  {journal} {Zh. Eksp. Teor. Fiz.}\ }\textbf {\bibinfo {volume} {53}},\ \bibinfo {pages} {1717} (\bibinfo {year} {1967})}\BibitemShut {NoStop}%
\bibitem [{\citenamefont {Mohapatra}\ \emph {et~al.}(2007)\citenamefont {Mohapatra} \emph {et~al.}}]{Mohapatra_2007}%
  \BibitemOpen
  \bibfield  {author} {\bibinfo {author} {\bibfnamefont {R.~N.}\ \bibnamefont {Mohapatra}} \emph {et~al.},\ }\bibfield  {title} {\bibinfo {title} {Theory of neutrinos: a white paper},\ }\href {https://doi.org/10.1088/0034-4885/70/11/R02} {\bibfield  {journal} {\bibinfo  {journal} {Rep. Prog.Phys.}\ }\textbf {\bibinfo {volume} {70}},\ \bibinfo {pages} {1757} (\bibinfo {year} {2007})}\BibitemShut {NoStop}%
\bibitem [{\citenamefont {Formaggio}\ \emph {et~al.}(2021)\citenamefont {Formaggio}, \citenamefont {de~Gouvea},\ and\ \citenamefont {Robertson}}]{formaggio2021direct}%
  \BibitemOpen
  \bibfield  {author} {\bibinfo {author} {\bibfnamefont {J.}~\bibnamefont {Formaggio}}, \bibinfo {author} {\bibfnamefont {A.}~\bibnamefont {de~Gouvea}},\ and\ \bibinfo {author} {\bibfnamefont {R.~G.~H.}\ \bibnamefont {Robertson}},\ }\bibfield  {title} {\bibinfo {title} {Direct measurements of neutrino mass},\ }\href {https://doi.org/10.1016/j.physrep.2021.02.002} {\bibfield  {journal} {\bibinfo  {journal} {Phys. Rep.}\ }\textbf {\bibinfo {volume} {914}},\ \bibinfo {pages} {1} (\bibinfo {year} {2021})}\BibitemShut {NoStop}%
\bibitem [{\citenamefont {Dolinski}\ \emph {et~al.}(2019)\citenamefont {Dolinski}, \citenamefont {Poon},\ and\ \citenamefont {Rodejohann}}]{dolinski2019neutrinoless}%
  \BibitemOpen
  \bibfield  {author} {\bibinfo {author} {\bibfnamefont {M.~J.}\ \bibnamefont {Dolinski}}, \bibinfo {author} {\bibfnamefont {A.~W.~P.}\ \bibnamefont {Poon}},\ and\ \bibinfo {author} {\bibfnamefont {W.}~\bibnamefont {Rodejohann}},\ }\bibfield  {title} {\bibinfo {title} {Neutrinoless double-beta decay: status and prospects},\ }\href {https://doi.org/10.1146/annurev-nucl-101918-023407} {\bibfield  {journal} {\bibinfo  {journal} {Annu. Rev. Nucl. Part. Sci.}\ }\textbf {\bibinfo {volume} {69}},\ \bibinfo {pages} {219} (\bibinfo {year} {2019})}\BibitemShut {NoStop}%
\bibitem [{\citenamefont {Hansen}\ \emph {et~al.}(2020)\citenamefont {Hansen}, \citenamefont {Lindner},\ and\ \citenamefont {Scholer}}]{hansen2020timing}%
  \BibitemOpen
  \bibfield  {author} {\bibinfo {author} {\bibfnamefont {R.~S.~L.}\ \bibnamefont {Hansen}}, \bibinfo {author} {\bibfnamefont {M.}~\bibnamefont {Lindner}},\ and\ \bibinfo {author} {\bibfnamefont {O.}~\bibnamefont {Scholer}},\ }\bibfield  {title} {\bibinfo {title} {Timing the neutrino signal of a galactic supernova},\ }\href {https://doi.org/10.1103/PhysRevD.101.123018} {\bibfield  {journal} {\bibinfo  {journal} {Phys. Rev. D}\ }\textbf {\bibinfo {volume} {101}},\ \bibinfo {pages} {123018} (\bibinfo {year} {2020})}\BibitemShut {NoStop}%
\bibitem [{\citenamefont {Horiuchi}\ and\ \citenamefont {Kneller}(2018)}]{horiuchi2018can}%
  \BibitemOpen
  \bibfield  {author} {\bibinfo {author} {\bibfnamefont {S.}~\bibnamefont {Horiuchi}}\ and\ \bibinfo {author} {\bibfnamefont {J.~P.}\ \bibnamefont {Kneller}},\ }\bibfield  {title} {\bibinfo {title} {What can be learned from a future supernova neutrino detection?},\ }\href {https://doi.org/10.1088/1361-6471/aaa90a} {\bibfield  {journal} {\bibinfo  {journal} {J. Phys. G}\ }\textbf {\bibinfo {volume} {45}},\ \bibinfo {pages} {043002} (\bibinfo {year} {2018})}\BibitemShut {NoStop}%
\bibitem [{\citenamefont {Lesgourgues}\ and\ \citenamefont {Pastor}(2012)}]{lesgourgues2012neutrino}%
  \BibitemOpen
  \bibfield  {author} {\bibinfo {author} {\bibfnamefont {J.}~\bibnamefont {Lesgourgues}}\ and\ \bibinfo {author} {\bibfnamefont {S.}~\bibnamefont {Pastor}},\ }\bibfield  {title} {\bibinfo {title} {Neutrino mass from cosmology},\ }\href {https://doi.org/10.1155/2012/608515} {\bibfield  {journal} {\bibinfo  {journal} {Adv. High Energy Phys.}\ }\textbf {\bibinfo {volume} {2012}},\ \bibinfo {pages} {608515} (\bibinfo {year} {2012})}\BibitemShut {NoStop}%
\bibitem [{\citenamefont {Beavis}\ \emph {et~al.}(1995)\citenamefont {Beavis}, \citenamefont {Carroll},\ and\ \citenamefont {Chiang}}]{beavis1995long}%
  \BibitemOpen
  \bibfield  {author} {\bibinfo {author} {\bibfnamefont {D.}~\bibnamefont {Beavis}}, \bibinfo {author} {\bibfnamefont {A.}~\bibnamefont {Carroll}},\ and\ \bibinfo {author} {\bibfnamefont {I.}~\bibnamefont {Chiang}},\ }\href {https://doi.org/10.2172/52878} {\emph {\bibinfo {title} {{Long baseline neutrino oscillation experiment at the AGS. Physics design report}}}},\ \bibinfo {type} {Tech. Rep.}\ \bibinfo {number} {52459}\ (\bibinfo  {institution} {Brookhaven National Lab.},\ \bibinfo {year} {1995})\BibitemShut {NoStop}%
\bibitem [{\citenamefont {Helmer}(1994)}]{Helmer:1994ac}%
  \BibitemOpen
  \bibfield  {author} {\bibinfo {author} {\bibfnamefont {R.~L.}\ \bibnamefont {Helmer}},\ }\bibfield  {title} {\bibinfo {title} {{A new long baseline neutrino oscillation experiment at Brookhaven}},\ }in\ \href@noop {} {\emph {\bibinfo {booktitle} {{9th Lake Louise Winter Institute: Particle Physics and Cosmology}}}}\ (\bibinfo {year} {1994})\ pp.\ \bibinfo {pages} {0291--301}\BibitemShut {NoStop}%
\bibitem [{\citenamefont {Abe}\ \emph {et~al.}(2011)\citenamefont {Abe} \emph {et~al.}}]{T2K:2011qtm}%
  \BibitemOpen
  \bibfield  {author} {\bibinfo {author} {\bibfnamefont {K.}~\bibnamefont {Abe}} \emph {et~al.} (\bibinfo {collaboration} {T2K}),\ }\bibfield  {title} {\bibinfo {title} {{The T2K experiment}},\ }\href {https://doi.org/10.1016/j.nima.2011.06.067} {\bibfield  {journal} {\bibinfo  {journal} {Nucl. Instrum. Meth. A}\ }\textbf {\bibinfo {volume} {659}},\ \bibinfo {pages} {106} (\bibinfo {year} {2011})}\BibitemShut {NoStop}%
\bibitem [{\citenamefont {Ayres}\ \emph {et~al.}(2007)\citenamefont {Ayres} \emph {et~al.}}]{NOvA:2007rmc}%
  \BibitemOpen
  \bibfield  {author} {\bibinfo {author} {\bibfnamefont {D.~S.}\ \bibnamefont {Ayres}} \emph {et~al.} (\bibinfo {collaboration} {NOvA}),\ }\href {https://doi.org/10.2172/935497} {\emph {\bibinfo {title} {{The NOvA technical design report}}}},\ \bibinfo {type} {Tech. Rep.}\ \bibinfo {number} {FERMILAB-DESIGN-2007-01}\ (\bibinfo  {institution} {Fermilab National Accelerator Laboratory},\ \bibinfo {year} {2007})\BibitemShut {NoStop}%
\bibitem [{\citenamefont {Di~Lodovico}\ \emph {et~al.}(2023)\citenamefont {Di~Lodovico}, \citenamefont {Patterson}, \citenamefont {Shiozawa},\ and\ \citenamefont {Worcester}}]{DiLodovico:2023jgr}%
  \BibitemOpen
  \bibfield  {author} {\bibinfo {author} {\bibfnamefont {F.}~\bibnamefont {Di~Lodovico}}, \bibinfo {author} {\bibfnamefont {R.~B.}\ \bibnamefont {Patterson}}, \bibinfo {author} {\bibfnamefont {M.}~\bibnamefont {Shiozawa}},\ and\ \bibinfo {author} {\bibfnamefont {E.}~\bibnamefont {Worcester}},\ }\bibfield  {title} {\bibinfo {title} {Experimental considerations in long-baseline neutrino oscillation measurements},\ }\href {https://doi.org/10.1146/annurev-nucl-102020-101615} {\bibfield  {journal} {\bibinfo  {journal} {Ann. Rev. Nucl. Part. Sci.}\ }\textbf {\bibinfo {volume} {73}},\ \bibinfo {pages} {69} (\bibinfo {year} {2023})}\BibitemShut {NoStop}%
\bibitem [{\citenamefont {Metropolis}\ \emph {et~al.}(1953)\citenamefont {Metropolis}, \citenamefont {Rosenbluth}, \citenamefont {Rosenbluth}, \citenamefont {Teller},\ and\ \citenamefont {Teller}}]{Metropolis:1953am}%
  \BibitemOpen
  \bibfield  {author} {\bibinfo {author} {\bibfnamefont {N.}~\bibnamefont {Metropolis}}, \bibinfo {author} {\bibfnamefont {A.~W.}\ \bibnamefont {Rosenbluth}}, \bibinfo {author} {\bibfnamefont {M.~N.}\ \bibnamefont {Rosenbluth}}, \bibinfo {author} {\bibfnamefont {A.~H.}\ \bibnamefont {Teller}},\ and\ \bibinfo {author} {\bibfnamefont {E.}~\bibnamefont {Teller}},\ }\bibfield  {title} {\bibinfo {title} {{Equation of state calculations by fast computing machines}},\ }\href {https://doi.org/10.1063/1.1699114} {\bibfield  {journal} {\bibinfo  {journal} {J. Chem. Phys.}\ }\textbf {\bibinfo {volume} {21}},\ \bibinfo {pages} {1087} (\bibinfo {year} {1953})}\BibitemShut {NoStop}%
\bibitem [{\citenamefont {Hastings}(1970)}]{Hastings:1970aa}%
  \BibitemOpen
  \bibfield  {author} {\bibinfo {author} {\bibfnamefont {W.~K.}\ \bibnamefont {Hastings}},\ }\bibfield  {title} {\bibinfo {title} {{Monte Carlo sampling methods using Markov chains and their applications}},\ }\href {https://doi.org/10.1093/biomet/57.1.97} {\bibfield  {journal} {\bibinfo  {journal} {Biometrika}\ }\textbf {\bibinfo {volume} {57}},\ \bibinfo {pages} {97} (\bibinfo {year} {1970})}\BibitemShut {NoStop}%
\bibitem [{\citenamefont {Acero}\ \emph {et~al.}(2024)\citenamefont {Acero} \emph {et~al.}}]{NOvA:2023iam}%
  \BibitemOpen
  \bibfield  {author} {\bibinfo {author} {\bibfnamefont {M.~A.}\ \bibnamefont {Acero}} \emph {et~al.} (\bibinfo {collaboration} {NOvA}),\ }\bibfield  {title} {\bibinfo {title} {{Expanding neutrino oscillation parameter measurements in NOvA using a Bayesian approach}},\ }\href {https://doi.org/10.1103/PhysRevD.110.012005} {\bibfield  {journal} {\bibinfo  {journal} {Phys. Rev. D}\ }\textbf {\bibinfo {volume} {110}},\ \bibinfo {pages} {012005} (\bibinfo {year} {2024})}\BibitemShut {NoStop}%
\bibitem [{\citenamefont {{The MaCh3 Collaboration}}(2024)}]{the_mach3_collaboration_2024}%
  \BibitemOpen
  \bibfield  {author} {\bibinfo {author} {\bibnamefont {{The MaCh3 Collaboration}}},\ }\href {https://doi.org/10.5281/zenodo.7608367} {\bibinfo {title} {{MaCh3}}} (\bibinfo {year} {2024})\BibitemShut {NoStop}%
\bibitem [{\citenamefont {Kurtzer}\ \emph {et~al.}(2017)\citenamefont {Kurtzer}, \citenamefont {Sochat},\ and\ \citenamefont {Bauer}}]{kurtzer2017singularity}%
  \BibitemOpen
  \bibfield  {author} {\bibinfo {author} {\bibfnamefont {G.~M.}\ \bibnamefont {Kurtzer}}, \bibinfo {author} {\bibfnamefont {V.}~\bibnamefont {Sochat}},\ and\ \bibinfo {author} {\bibfnamefont {M.~W.}\ \bibnamefont {Bauer}},\ }\bibfield  {title} {\bibinfo {title} {Singularity: Scientific containers for mobility of compute},\ }\href {https://doi.org/10.1371/journal.pone.0177459} {\bibfield  {journal} {\bibinfo  {journal} {PloS one}\ }\textbf {\bibinfo {volume} {12}},\ \bibinfo {pages} {e0177459} (\bibinfo {year} {2017})}\BibitemShut {NoStop}%
\bibitem [{\citenamefont {Alt}\ \emph {et~al.}(2007)\citenamefont {Alt} \emph {et~al.}}]{NA49:2006oyk}%
  \BibitemOpen
  \bibfield  {author} {\bibinfo {author} {\bibfnamefont {C.}~\bibnamefont {Alt}} \emph {et~al.} (\bibinfo {collaboration} {NA49}),\ }\bibfield  {title} {\bibinfo {title} {{Inclusive production of charged pions in p+C collisions at 158-GeV/c beam momentum}},\ }\href {https://doi.org/10.1140/epjc/s10052-006-0165-7} {\bibfield  {journal} {\bibinfo  {journal} {Eur. Phys. J. C}\ }\textbf {\bibinfo {volume} {49}},\ \bibinfo {pages} {897} (\bibinfo {year} {2007})}\BibitemShut {NoStop}%
\bibitem [{\citenamefont {Abgrall}\ \emph {et~al.}(2016{\natexlab{a}})\citenamefont {Abgrall} \emph {et~al.}}]{NA61SHINE:2016nlf}%
  \BibitemOpen
  \bibfield  {author} {\bibinfo {author} {\bibfnamefont {N.}~\bibnamefont {Abgrall}} \emph {et~al.} (\bibinfo {collaboration} {NA61/SHINE}),\ }\bibfield  {title} {\bibinfo {title} {{Measurements of $\pi ^\pm $ differential yields from the surface of the T2K replica target for incoming 31 GeV/c protons with the NA61/SHINE spectrometer at the CERN SPS}},\ }\href {https://doi.org/10.1140/epjc/s10052-016-4440-y} {\bibfield  {journal} {\bibinfo  {journal} {Eur. Phys. J. C}\ }\textbf {\bibinfo {volume} {76}},\ \bibinfo {pages} {617} (\bibinfo {year} {2016}{\natexlab{a}})}\BibitemShut {NoStop}%
\bibitem [{\citenamefont {Abgrall}\ \emph {et~al.}(2016{\natexlab{b}})\citenamefont {Abgrall} \emph {et~al.}}]{NA61SHINE:2015bad}%
  \BibitemOpen
  \bibfield  {author} {\bibinfo {author} {\bibfnamefont {N.}~\bibnamefont {Abgrall}} \emph {et~al.} (\bibinfo {collaboration} {NA61/SHINE}),\ }\bibfield  {title} {\bibinfo {title} {{Measurements of $\pi ^{\pm }$ , $K^{\pm }$ , $K^0_S$ , $\varLambda $ and proton production in proton\textendash{}carbon interactions at 31 GeV/c with the NA61/SHINE spectrometer at the CERN SPS}},\ }\href {https://doi.org/10.1140/epjc/s10052-016-3898-y} {\bibfield  {journal} {\bibinfo  {journal} {Eur. Phys. J. C}\ }\textbf {\bibinfo {volume} {76}},\ \bibinfo {pages} {84} (\bibinfo {year} {2016}{\natexlab{b}})}\BibitemShut {NoStop}%
\bibitem [{\citenamefont {Hayato}\ and\ \citenamefont {Pickering}(2021)}]{Hayato:2021heg}%
  \BibitemOpen
  \bibfield  {author} {\bibinfo {author} {\bibfnamefont {Y.}~\bibnamefont {Hayato}}\ and\ \bibinfo {author} {\bibfnamefont {L.}~\bibnamefont {Pickering}},\ }\bibfield  {title} {\bibinfo {title} {{The NEUT neutrino interaction simulation program library}},\ }\href {https://doi.org/10.1140/epjs/s11734-021-00287-7} {\bibfield  {journal} {\bibinfo  {journal} {Eur. Phys. J. ST}\ }\textbf {\bibinfo {volume} {230}},\ \bibinfo {pages} {4469} (\bibinfo {year} {2021})}\BibitemShut {NoStop}%
\bibitem [{\citenamefont {Andreopoulos}\ \emph {et~al.}(2010)\citenamefont {Andreopoulos} \emph {et~al.}}]{Andreopoulos:2009rq}%
  \BibitemOpen
  \bibfield  {author} {\bibinfo {author} {\bibfnamefont {C.}~\bibnamefont {Andreopoulos}} \emph {et~al.},\ }\bibfield  {title} {\bibinfo {title} {{The GENIE neutrino Monte Carlo generator}},\ }\href {https://doi.org/10.1016/j.nima.2009.12.009} {\bibfield  {journal} {\bibinfo  {journal} {Nucl. Instrum. Meth. A}\ }\textbf {\bibinfo {volume} {614}},\ \bibinfo {pages} {87} (\bibinfo {year} {2010})}\BibitemShut {NoStop}%
\bibitem [{\citenamefont {Balantekin}\ \emph {et~al.}(2022)\citenamefont {Balantekin} \emph {et~al.}}]{Balantekin:2022jrq}%
  \BibitemOpen
  \bibfield  {author} {\bibinfo {author} {\bibfnamefont {A.~B.}\ \bibnamefont {Balantekin}} \emph {et~al.},\ }\bibfield  {title} {\bibinfo {title} {{Snowmass Neutrino Frontier: Neutrino Interaction Cross Sections (NF06) Topical Group Report}}\ }(\bibinfo {year} {2022})\ \Eprint {https://arxiv.org/abs/2209.06872} {arXiv:2209.06872 [hep-ex]} \BibitemShut {NoStop}%
\bibitem [{\citenamefont {Day}\ and\ \citenamefont {McFarland}(2012)}]{day2012differences}%
  \BibitemOpen
  \bibfield  {author} {\bibinfo {author} {\bibfnamefont {M.}~\bibnamefont {Day}}\ and\ \bibinfo {author} {\bibfnamefont {K.~S.}\ \bibnamefont {McFarland}},\ }\bibfield  {title} {\bibinfo {title} {Differences in quasielastic cross sections of muon and electron neutrinos},\ }\href {https://doi.org/10.1103/PhysRevD.86.053003} {\bibfield  {journal} {\bibinfo  {journal} {Phys. Rev. D}\ }\textbf {\bibinfo {volume} {86}},\ \bibinfo {pages} {053003} (\bibinfo {year} {2012})}\BibitemShut {NoStop}%
\bibitem [{\citenamefont {Zyla}\ \emph {et~al.}(2020)\citenamefont {Zyla} \emph {et~al.}}]{ParticleDataGroup:2020ssz}%
  \BibitemOpen
  \bibfield  {author} {\bibinfo {author} {\bibfnamefont {P.~A.}\ \bibnamefont {Zyla}} \emph {et~al.} (\bibinfo {collaboration} {Particle Data Group}),\ }\bibfield  {title} {\bibinfo {title} {{Review of particle physics}},\ }\href {https://doi.org/10.1093/ptep/ptaa104} {\bibfield  {journal} {\bibinfo  {journal} {PTEP}\ }\textbf {\bibinfo {volume} {2020}},\ \bibinfo {pages} {083C01} (\bibinfo {year} {2020})}\BibitemShut {NoStop}%
\bibitem [{\citenamefont {Gelman}\ \emph {et~al.}(1996)\citenamefont {Gelman}, \citenamefont {Meng},\ and\ \citenamefont {Stern}}]{gellmanpospred}%
  \BibitemOpen
  \bibfield  {author} {\bibinfo {author} {\bibfnamefont {A.}~\bibnamefont {Gelman}}, \bibinfo {author} {\bibfnamefont {X.}~\bibnamefont {Meng}},\ and\ \bibinfo {author} {\bibfnamefont {H.}~\bibnamefont {Stern}},\ }\bibfield  {title} {\bibinfo {title} {Posterior predictive assessment of model fitness via realized discrepancies},\ }\href {http://www.jstor.org/stable/24306036} {\bibfield  {journal} {\bibinfo  {journal} {Stat. Sin.}\ }\textbf {\bibinfo {volume} {6}},\ \bibinfo {pages} {733} (\bibinfo {year} {1996})}\BibitemShut {NoStop}%
\bibitem [{\citenamefont {Adamson}\ \emph {et~al.}(2020)\citenamefont {Adamson} \emph {et~al.}}]{MINOS:2020llm}%
  \BibitemOpen
  \bibfield  {author} {\bibinfo {author} {\bibfnamefont {P.}~\bibnamefont {Adamson}} \emph {et~al.} (\bibinfo {collaboration} {MINOS+}),\ }\bibfield  {title} {\bibinfo {title} {{Precision constraints for three-flavor neutrino oscillations from the full MINOS+ and MINOS dataset}},\ }\href {https://doi.org/10.1103/PhysRevLett.125.131802} {\bibfield  {journal} {\bibinfo  {journal} {Phys. Rev. Lett.}\ }\textbf {\bibinfo {volume} {125}},\ \bibinfo {pages} {131802} (\bibinfo {year} {2020})}\BibitemShut {NoStop}%
\bibitem [{\citenamefont {Abbasi}\ \emph {et~al.}(2025)\citenamefont {Abbasi} \emph {et~al.}}]{IceCubeCollaboration:2024zec}%
  \BibitemOpen
  \bibfield  {author} {\bibinfo {author} {\bibfnamefont {R.}~\bibnamefont {Abbasi}} \emph {et~al.} (\bibinfo {collaboration} {IceCube Collaboration}),\ }\bibfield  {title} {\bibinfo {title} {{Measurement of Atmospheric Neutrino Oscillation Parameters Using Convolutional Neural Networks with 9.3 Years of Data in IceCube DeepCore}},\ }\href {https://doi.org/10.1103/PhysRevLett.134.091801} {\bibfield  {journal} {\bibinfo  {journal} {Phys. Rev. Lett.}\ }\textbf {\bibinfo {volume} {134}},\ \bibinfo {pages} {091801} (\bibinfo {year} {2025})}\BibitemShut {NoStop}%
\bibitem [{\citenamefont {Abe}\ \emph {et~al.}(2025)\citenamefont {Abe} \emph {et~al.}}]{abe2024first}%
  \BibitemOpen
  \bibfield  {author} {\bibinfo {author} {\bibfnamefont {K.}~\bibnamefont {Abe}} \emph {et~al.} (\bibinfo {collaboration} {Super-Kamiokande Collaboration and T2K Collaboration}),\ }\bibfield  {title} {\bibinfo {title} {{First joint oscillation analysis of Super-Kamiokande atmospheric and T2K accelerator neutrino data}},\ }\href {https://doi.org/10.1103/PhysRevLett.134.011801} {\bibfield  {journal} {\bibinfo  {journal} {Phys. Rev. Lett.}\ }\textbf {\bibinfo {volume} {134}},\ \bibinfo {pages} {011801} (\bibinfo {year} {2025})}\BibitemShut {NoStop}%
\bibitem [{\citenamefont {Wester}\ \emph {et~al.}(2024)\citenamefont {Wester} \emph {et~al.}}]{Super-Kamiokande:2023ahc}%
  \BibitemOpen
  \bibfield  {author} {\bibinfo {author} {\bibfnamefont {T.}~\bibnamefont {Wester}} \emph {et~al.} (\bibinfo {collaboration} {The Super-Kamiokande Collaboration}),\ }\bibfield  {title} {\bibinfo {title} {{Atmospheric neutrino oscillation analysis with neutron tagging and an expanded fiducial volume in Super-Kamiokande I--V}},\ }\href {https://doi.org/10.1103/PhysRevD.109.072014} {\bibfield  {journal} {\bibinfo  {journal} {Phys. Rev. D}\ }\textbf {\bibinfo {volume} {109}},\ \bibinfo {pages} {072014} (\bibinfo {year} {2024})}\BibitemShut {NoStop}%
\bibitem [{\citenamefont {An}\ \emph {et~al.}(2023)\citenamefont {An} \emph {et~al.}}]{DayaBay:2022orm}%
  \BibitemOpen
  \bibfield  {author} {\bibinfo {author} {\bibfnamefont {F.~P.}\ \bibnamefont {An}} \emph {et~al.} (\bibinfo {collaboration} {Daya Bay}),\ }\bibfield  {title} {\bibinfo {title} {{Precision measurement of reactor antineutrino oscillation at kilometer-scale baselines by Daya Bay}},\ }\href {https://doi.org/10.1103/PhysRevLett.130.161802} {\bibfield  {journal} {\bibinfo  {journal} {Phys. Rev. Lett.}\ }\textbf {\bibinfo {volume} {130}},\ \bibinfo {pages} {161802} (\bibinfo {year} {2023})}\BibitemShut {NoStop}%
\bibitem [{\citenamefont {Jeon}\ \emph {et~al.}(2025)\citenamefont {Jeon} \emph {et~al.}}]{RENO:2024msr}%
  \BibitemOpen
  \bibfield  {author} {\bibinfo {author} {\bibfnamefont {S.}~\bibnamefont {Jeon}} \emph {et~al.} (\bibinfo {collaboration} {RENO}),\ }\bibfield  {title} {\bibinfo {title} {{Measurement of reactor antineutrino oscillation amplitude and frequency using 3800 days of complete data sample of the RENO experiment}},\ }\href {https://doi.org/10.1103/dc6j-5ky6} {\bibfield  {journal} {\bibinfo  {journal} {Phys. Rev. D}\ }\textbf {\bibinfo {volume} {111}},\ \bibinfo {pages} {112006} (\bibinfo {year} {2025})}\BibitemShut {NoStop}%
\bibitem [{\citenamefont {An}\ \emph {et~al.}(2024)\citenamefont {An} \emph {et~al.}}]{DayaBay:2024hrv}%
  \BibitemOpen
  \bibfield  {author} {\bibinfo {author} {\bibfnamefont {F.~P.}\ \bibnamefont {An}} \emph {et~al.} (\bibinfo {collaboration} {Daya Bay}),\ }\bibfield  {title} {\bibinfo {title} {{Measurement of Electron Antineutrino Oscillation Amplitude and Frequency via Neutron Capture on Hydrogen at Daya Bay}},\ }\href {https://doi.org/10.1103/PhysRevLett.133.151801} {\bibfield  {journal} {\bibinfo  {journal} {Phys. Rev. Lett.}\ }\textbf {\bibinfo {volume} {133}},\ \bibinfo {pages} {151801} (\bibinfo {year} {2024})}\BibitemShut {NoStop}%
\bibitem [{\citenamefont {Jarlskog}(1985)}]{jarlskog}%
  \BibitemOpen
  \bibfield  {author} {\bibinfo {author} {\bibfnamefont {C.}~\bibnamefont {Jarlskog}},\ }\bibfield  {title} {\bibinfo {title} {{A basis independent formulation of the connection between quark mass matrices, CP violation and experiment}},\ }\href@noop {} {\bibfield  {journal} {\bibinfo  {journal} {Z. Phys. C}\ }\textbf {\bibinfo {volume} {29}},\ \bibinfo {pages} {491} (\bibinfo {year} {1985})}\BibitemShut {NoStop}%
\bibitem [{\citenamefont {Adamson}\ \emph {et~al.}(2016)\citenamefont {Adamson} \emph {et~al.}}]{Adamson:2015dkw}%
  \BibitemOpen
  \bibfield  {author} {\bibinfo {author} {\bibfnamefont {P.}~\bibnamefont {Adamson}} \emph {et~al.},\ }\bibfield  {title} {\bibinfo {title} {{The NuMI Neutrino Beam}},\ }\href {https://doi.org/10.1016/j.nima.2015.08.063} {\bibfield  {journal} {\bibinfo  {journal} {Nucl. Instrum. Meth. A}\ }\textbf {\bibinfo {volume} {806}},\ \bibinfo {pages} {279} (\bibinfo {year} {2016})}\BibitemShut {NoStop}%
\bibitem [{\citenamefont {Aurisano}\ \emph {et~al.}(2016)\citenamefont {Aurisano} \emph {et~al.}}]{Aurisano:2016jvx}%
  \BibitemOpen
  \bibfield  {author} {\bibinfo {author} {\bibfnamefont {A.}~\bibnamefont {Aurisano}} \emph {et~al.},\ }\bibfield  {title} {\bibinfo {title} {A convolutional neural network neutrino event classifier},\ }\href {https://doi.org/10.1088/1748-0221/11/09/P09001} {\bibfield  {journal} {\bibinfo  {journal} {JINST}\ }\textbf {\bibinfo {volume} {11}}\bibinfo  {number} { (09)},\ \bibinfo {pages} {P09001}}\BibitemShut {NoStop}%
\bibitem [{\citenamefont {Abe}\ \emph {et~al.}(2013)\citenamefont {Abe} \emph {et~al.}}]{T2K:2012bge}%
  \BibitemOpen
\bibfield  {number} {  }\bibfield  {author} {\bibinfo {author} {\bibfnamefont {K.}~\bibnamefont {Abe}} \emph {et~al.} (\bibinfo {collaboration} {T2K}),\ }\bibfield  {title} {\bibinfo {title} {{T2K neutrino flux prediction}},\ }\href {https://doi.org/10.1103/PhysRevD.87.012001} {\bibfield  {journal} {\bibinfo  {journal} {Phys. Rev. D}\ }\textbf {\bibinfo {volume} {87}},\ \bibinfo {pages} {012001} (\bibinfo {year} {2013})},\ \bibinfo {note} {[Addendum: Phys.Rev.D 87, 019902 (2013)]}\BibitemShut {NoStop}%
\bibitem [{\citenamefont {Aliaga}\ \emph {et~al.}(2016)\citenamefont {Aliaga} \emph {et~al.}}]{MINERvA:2016iqn}%
  \BibitemOpen
  \bibfield  {author} {\bibinfo {author} {\bibfnamefont {L.}~\bibnamefont {Aliaga}} \emph {et~al.} (\bibinfo {collaboration} {MINERvA}),\ }\bibfield  {title} {\bibinfo {title} {{Neutrino Flux Predictions for the NuMI Beam}},\ }\href {https://doi.org/10.1103/PhysRevD.94.092005} {\bibfield  {journal} {\bibinfo  {journal} {Phys. Rev. D}\ }\textbf {\bibinfo {volume} {94}},\ \bibinfo {pages} {092005} (\bibinfo {year} {2016})},\ \bibinfo {note} {[Addendum: Phys.Rev.D 95, 039903 (2017)]}\BibitemShut {NoStop}%
\bibitem [{\citenamefont {Esteban}\ \emph {et~al.}(2020)\citenamefont {Esteban}, \citenamefont {Gonzalez-Garcia}, \citenamefont {Maltoni}, \citenamefont {Schwetz},\ and\ \citenamefont {Zhou}}]{Esteban:2020cvm}%
  \BibitemOpen
  \bibfield  {author} {\bibinfo {author} {\bibfnamefont {I.}~\bibnamefont {Esteban}}, \bibinfo {author} {\bibfnamefont {M.~C.}\ \bibnamefont {Gonzalez-Garcia}}, \bibinfo {author} {\bibfnamefont {M.}~\bibnamefont {Maltoni}}, \bibinfo {author} {\bibfnamefont {T.}~\bibnamefont {Schwetz}},\ and\ \bibinfo {author} {\bibfnamefont {A.}~\bibnamefont {Zhou}},\ }\bibfield  {title} {\bibinfo {title} {{The fate of hints: updated global analysis of three-flavor neutrino oscillations}},\ }\href {https://doi.org/10.1007/JHEP09(2020)178} {\bibfield  {journal} {\bibinfo  {journal} {J. High Energy Phys.}\ }\textbf {\bibinfo {volume} {2020}}\bibinfo  {number} { (178)}}\BibitemShut {NoStop}%
\bibitem [{\citenamefont {Stowell}\ \emph {et~al.}(2019)\citenamefont {Stowell} \emph {et~al.}}]{MINERvA:2019kfr}%
  \BibitemOpen
\bibfield  {number} {  }\bibfield  {author} {\bibinfo {author} {\bibfnamefont {P.}~\bibnamefont {Stowell}} \emph {et~al.} (\bibinfo {collaboration} {MINERvA}),\ }\bibfield  {title} {\bibinfo {title} {{Tuning the GENIE Pion Production Model with MINER$\nu$A Data}},\ }\href {https://doi.org/10.1103/PhysRevD.100.072005} {\bibfield  {journal} {\bibinfo  {journal} {Phys. Rev. D}\ }\textbf {\bibinfo {volume} {100}},\ \bibinfo {pages} {072005} (\bibinfo {year} {2019})}\BibitemShut {NoStop}%
\bibitem [{\citenamefont {Agostinelli}\ \emph {et~al.}(2003)\citenamefont {Agostinelli} \emph {et~al.}}]{GEANT4:2002zbu}%
  \BibitemOpen
  \bibfield  {author} {\bibinfo {author} {\bibfnamefont {S.}~\bibnamefont {Agostinelli}} \emph {et~al.} (\bibinfo {collaboration} {GEANT4}),\ }\bibfield  {title} {\bibinfo {title} {{GEANT4--a simulation toolkit}},\ }\href {https://doi.org/10.1016/S0168-9002(03)01368-8} {\bibfield  {journal} {\bibinfo  {journal} {Nucl. Instrum. Meth. A}\ }\textbf {\bibinfo {volume} {506}},\ \bibinfo {pages} {250} (\bibinfo {year} {2003})}\BibitemShut {NoStop}%
\bibitem [{\citenamefont {Salcedo}\ \emph {et~al.}(1988)\citenamefont {Salcedo}, \citenamefont {Oset}, \citenamefont {Vicente-Vacas},\ and\ \citenamefont {Garcia-Recio}}]{Salcedo:1987md}%
  \BibitemOpen
  \bibfield  {author} {\bibinfo {author} {\bibfnamefont {L.~L.}\ \bibnamefont {Salcedo}}, \bibinfo {author} {\bibfnamefont {E.}~\bibnamefont {Oset}}, \bibinfo {author} {\bibfnamefont {M.~J.}\ \bibnamefont {Vicente-Vacas}},\ and\ \bibinfo {author} {\bibfnamefont {C.}~\bibnamefont {Garcia-Recio}},\ }\bibfield  {title} {\bibinfo {title} {{Computer simulation of inclusive pion nuclear reactions}},\ }\href {https://doi.org/10.1016/0375-9474(88)90310-7} {\bibfield  {journal} {\bibinfo  {journal} {Nucl. Phys. A}\ }\textbf {\bibinfo {volume} {484}},\ \bibinfo {pages} {557} (\bibinfo {year} {1988})}\BibitemShut {NoStop}%
\bibitem [{\citenamefont {Guerra}\ \emph {et~al.}(2019)\citenamefont {Guerra} \emph {et~al.}}]{PhysRevD.99.052007}%
  \BibitemOpen
  \bibfield  {author} {\bibinfo {author} {\bibfnamefont {E.~S.~P.}\ \bibnamefont {Guerra}} \emph {et~al.},\ }\bibfield  {title} {\bibinfo {title} {Using world ${\ensuremath{\pi}}^{\ifmmode\pm\else\textpm\fi{}}$-nucleus scattering data to constrain an intranuclear cascade model},\ }\href {https://doi.org/10.1103/PhysRevD.99.052007} {\bibfield  {journal} {\bibinfo  {journal} {Phys. Rev. D}\ }\textbf {\bibinfo {volume} {99}},\ \bibinfo {pages} {052007} (\bibinfo {year} {2019})}\BibitemShut {NoStop}%
\bibitem [{\citenamefont {Pandey}\ \emph {et~al.}(2015)\citenamefont {Pandey}, \citenamefont {Jachowicz}, \citenamefont {Van~Cuyck}, \citenamefont {Ryckebusch},\ and\ \citenamefont {Martini}}]{pandey2015low}%
  \BibitemOpen
  \bibfield  {author} {\bibinfo {author} {\bibfnamefont {V.}~\bibnamefont {Pandey}}, \bibinfo {author} {\bibfnamefont {N.}~\bibnamefont {Jachowicz}}, \bibinfo {author} {\bibfnamefont {T.}~\bibnamefont {Van~Cuyck}}, \bibinfo {author} {\bibfnamefont {J.}~\bibnamefont {Ryckebusch}},\ and\ \bibinfo {author} {\bibfnamefont {M.}~\bibnamefont {Martini}},\ }\bibfield  {title} {\bibinfo {title} {Low-energy excitations and quasielastic contribution to electron-nucleus and neutrino-nucleus scattering in the continuum random-phase approximation},\ }\href {https://doi.org/10.1103/PhysRevC.92.024606} {\bibfield  {journal} {\bibinfo  {journal} {Phys. Rev. C}\ }\textbf {\bibinfo {volume} {92}},\ \bibinfo {pages} {024606} (\bibinfo {year} {2015})}\BibitemShut {NoStop}%
\bibitem [{\citenamefont {Dunn}(1961)}]{Dunn01031961}%
  \BibitemOpen
  \bibfield  {author} {\bibinfo {author} {\bibfnamefont {O.~J.}\ \bibnamefont {Dunn}},\ }\bibfield  {title} {\bibinfo {title} {Multiple comparisons among means},\ }\href {https://doi.org/10.1080/01621459.1961.10482090} {\bibfield  {journal} {\bibinfo  {journal} {J. Am. Stat. Assoc.}\ }\textbf {\bibinfo {volume} {56}},\ \bibinfo {pages} {52} (\bibinfo {year} {1961})}\BibitemShut {NoStop}%
\bibitem [{\citenamefont {Gelman}\ \emph {et~al.}(2013)\citenamefont {Gelman}, \citenamefont {Carlin}, \citenamefont {Stern}, \citenamefont {Dunson}, \citenamefont {Vehtari},\ and\ \citenamefont {Rubin}}]{gelman_bayesian_2013}%
  \BibitemOpen
  \bibfield  {author} {\bibinfo {author} {\bibfnamefont {A.}~\bibnamefont {Gelman}}, \bibinfo {author} {\bibfnamefont {J.~B.}\ \bibnamefont {Carlin}}, \bibinfo {author} {\bibfnamefont {H.~S.}\ \bibnamefont {Stern}}, \bibinfo {author} {\bibfnamefont {D.~B.}\ \bibnamefont {Dunson}}, \bibinfo {author} {\bibfnamefont {A.}~\bibnamefont {Vehtari}},\ and\ \bibinfo {author} {\bibfnamefont {D.~B.}\ \bibnamefont {Rubin}},\ }\href@noop {} {\emph {\bibinfo {title} {Bayesian Data Analysis, Third Edition}}}\ (\bibinfo  {publisher} {CRC Press},\ \bibinfo {year} {2013})\BibitemShut {NoStop}%
\bibitem [{\citenamefont {Parke}\ and\ \citenamefont {Zukanovich-Funchal}(2025)}]{Parke:2024xre}%
  \BibitemOpen
  \bibfield  {author} {\bibinfo {author} {\bibfnamefont {S.~J.}\ \bibnamefont {Parke}}\ and\ \bibinfo {author} {\bibfnamefont {R.}~\bibnamefont {Zukanovich-Funchal}},\ }\bibfield  {title} {\bibinfo {title} {{Mass ordering sum rule for the neutrino disappearance channels in T2K, NOvA, and JUNO}},\ }\href {https://doi.org/10.1103/PhysRevD.111.013008} {\bibfield  {journal} {\bibinfo  {journal} {Phys. Rev. D}\ }\textbf {\bibinfo {volume} {111}},\ \bibinfo {pages} {013008} (\bibinfo {year} {2025})},\ \Eprint {https://arxiv.org/abs/2404.08733} {arXiv:2404.08733 [hep-ph]} \BibitemShut {NoStop}%
\bibitem [{\citenamefont {de~Kerret}\ \emph {et~al.}(2020)\citenamefont {de~Kerret} \emph {et~al.}}]{DoubleChooz:2019qbj}%
  \BibitemOpen
  \bibfield  {author} {\bibinfo {author} {\bibfnamefont {H.}~\bibnamefont {de~Kerret}} \emph {et~al.} (\bibinfo {collaboration} {Double Chooz}),\ }\bibfield  {title} {\bibinfo {title} {{Double Chooz $\theta_{13}$ measurement via total neutron capture detection}},\ }\href {https://doi.org/10.1038/s41567-020-0831-y} {\bibfield  {journal} {\bibinfo  {journal} {Nature Phys.}\ }\textbf {\bibinfo {volume} {16}},\ \bibinfo {pages} {558} (\bibinfo {year} {2020})}\BibitemShut {NoStop}%
\end{thebibliography}%

\end{document}